%% file: main.tex
\documentclass[preprint,12pt]{elsarticle}

\usepackage{graphicx}
\usepackage{amssymb}

\usepackage{multirow}

\usepackage{color}

\newcommand{\GdSOw}{$\rm Gd_2(\rm SO_4)_3\cdot \rm 8H_2O$\ }
\newcommand{\GdSO}{$\rm Gd_2(\rm SO_4)_3$\ }

\journal{Nucl. Instrum. Meth. A}

\begin{document}

\begin{frontmatter}


\title{First Gadolinium Loading to Super-Kamiokande}




\input{Authors}


\begin{abstract} 
In order to improve Super-Kamiokande's neutron detection efficiency and to thereby increase its sensitivity to the diffuse supernova neutrino background flux, 13~tons of \GdSOw (gadolinium sulfate octahydrate) was dissolved into the detector's otherwise ultrapure water from July 14 to August 17, 2020, marking the start of the SK-Gd phase of operations.
During the loading, water was continuously recirculated at a rate of 60 
m$^3$/h, extracting water from the top of the detector and mixing it with concentrated \GdSOw solution to create a 0.02\% solution of the Gd compound before injecting it into the bottom of the detector. 
A clear boundary between the Gd-loaded and pure water was maintained through the loading, enabling monitoring of the loading itself and the spatial uniformity of the Gd concentration over the 35 days it took to reach the top of the detector.
During the subsequent commissioning the recirculation rate was increased to 120~m$^3$/h, resulting in a constant and uniform distribution of Gd throughout the detector and water transparency equivalent to that of previous pure-water operation periods.
Using an Am-Be neutron calibration source the mean neutron capture time was measured to be $115\pm1$~$\mu$s, which corresponds to a Gd concentration of $111\pm2$~ppm, as expected for this level of Gd loading. 
This paper describes changes made to the water circulation system for this detector upgrade, the Gd loading procedure, detector commissioning, and the first neutron calibration measurements in SK-Gd.




\end{abstract}


\end{frontmatter}


\section{Introduction} 
\label{S:1}
The 50~kiloton water Cherenkov detector Super-Kamiokande (SK) has been operating for more than 25~years and continues to provide measurements and constraints covering a variety of topics in particle physics~\cite{pdacay:2020,nature:2020} and astroparticle physics~\cite{wimp:2020}. 
In order to extend the reach of SK to the as-yet unobserved diffuse supernova neutrino background (DSNB) flux, sometimes also called the supernova relic neutrino (SRN) flux, as well as improve SK's performance during a supernova explosion in our own galaxy, the authors of Ref.~\cite{beacom:2004} proposed adding gadolinium (Gd) to the detector's water.
As a result of its large neutron capture cross section and subsequent 8~MeV $\gamma$-ray cascade, the presence of Gd enhances the detector's ability to identify and differentiate antineutrino-induced events from backgrounds via the coincidence detection of neutrons from inverse beta decay reactions: $\bar{\nu}_{e(\mu)} + p \rightarrow e^+(\mu^+) + n$ . 
Although the original proposal was primarily motivated by relic neutrino observations, this technique was also intended to extend to higher energy phenomena, allowing for neutrino-antineutrino discrimination in the atmospheric and beam neutrino fluxes as well as improved background rejection in proton decay searches.

After successful development of the technology using a 200~ton water Cherenkov demonstrator called EGADS~\cite{egads:2020}, the Super-Kamiokande Collaboration initiated the SK-Gd project in 2015 with a final goal of adding 0.2\% \GdSOw (gadolinium sulfate octahydrate) by mass to the detector water.
Following further research and development to scale the technology to the 50~kiloton SK, the first step of SK-Gd began with the loading of 13~tons of $\rm Gd_2(\rm SO_4)_3\cdot \rm 8H_2O$, roughly 10\% of the final target concentration,  from July 14 to August 17, 2020.

In this paper we report the details of the Gd-loading and water circulation system, methods to control the flow of water in the detector, as well as calibration measurements to confirm the loaded Gd concentration.
In Section 2 we describe the SK-Gd water system and its operation scheme (detailed specifications of the water system are described in Appendix A) before presenting details of the flow control method in Section 3. 
These can be compared with the flow configuration of previous detector phases described in Appendix B.
The Gd loading for this first stage of SK-Gd is described in Section 4.
Future prospects and concluding remarks are presented in Section 5.

\section{SK-Gd Water System} 
\label{S:2}
\begin{figure}[htb!]
\centering\includegraphics[width=1.0\linewidth, trim = 150 0 150 0]{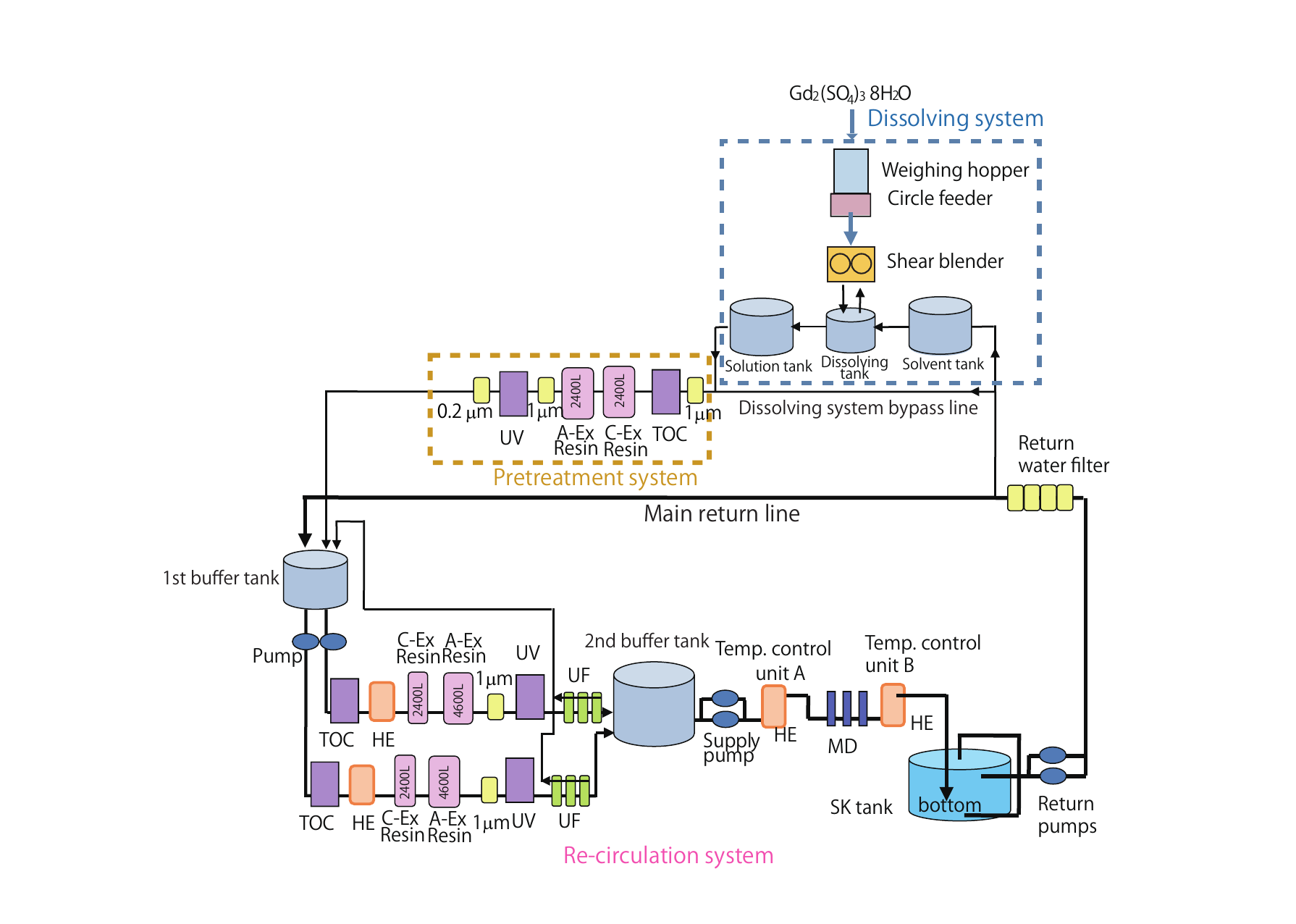}
\caption{Schematic diagram of SK-Gd water system.}
\label{fig:gd-water-sys}
\end{figure}

With the exception of a few elements inherited from the SK ultrapure water system, the SK-Gd water system was newly designed and constructed to dissolve \GdSOw into the detector water, pretreat it to remove impurities, and then to circulate and continuously purify the resulting 50 kilotons of Gd-loaded water. 
The new system purifies the water in the same way as the previous system, removing contaminants such as bacteria, particulates, and most dissolved ions, while simultaneously preserving the dissolved gadolinium (Gd$^{3+}$) and sulfate (SO$_4^{2-}$) ions in solution.
Water from the detector was continuously circulated and purified before, during, and after the Gd loading. 
A schematic diagram of the SK-Gd water system is shown in Fig.~\ref{fig:gd-water-sys} and described below.
Detailed specifications for each element in the system are listed in \ref{Append:1}.

\subsection{Gd-dissolving System}

\begin{figure}[htb]
\centering\includegraphics[width=0.8\linewidth]{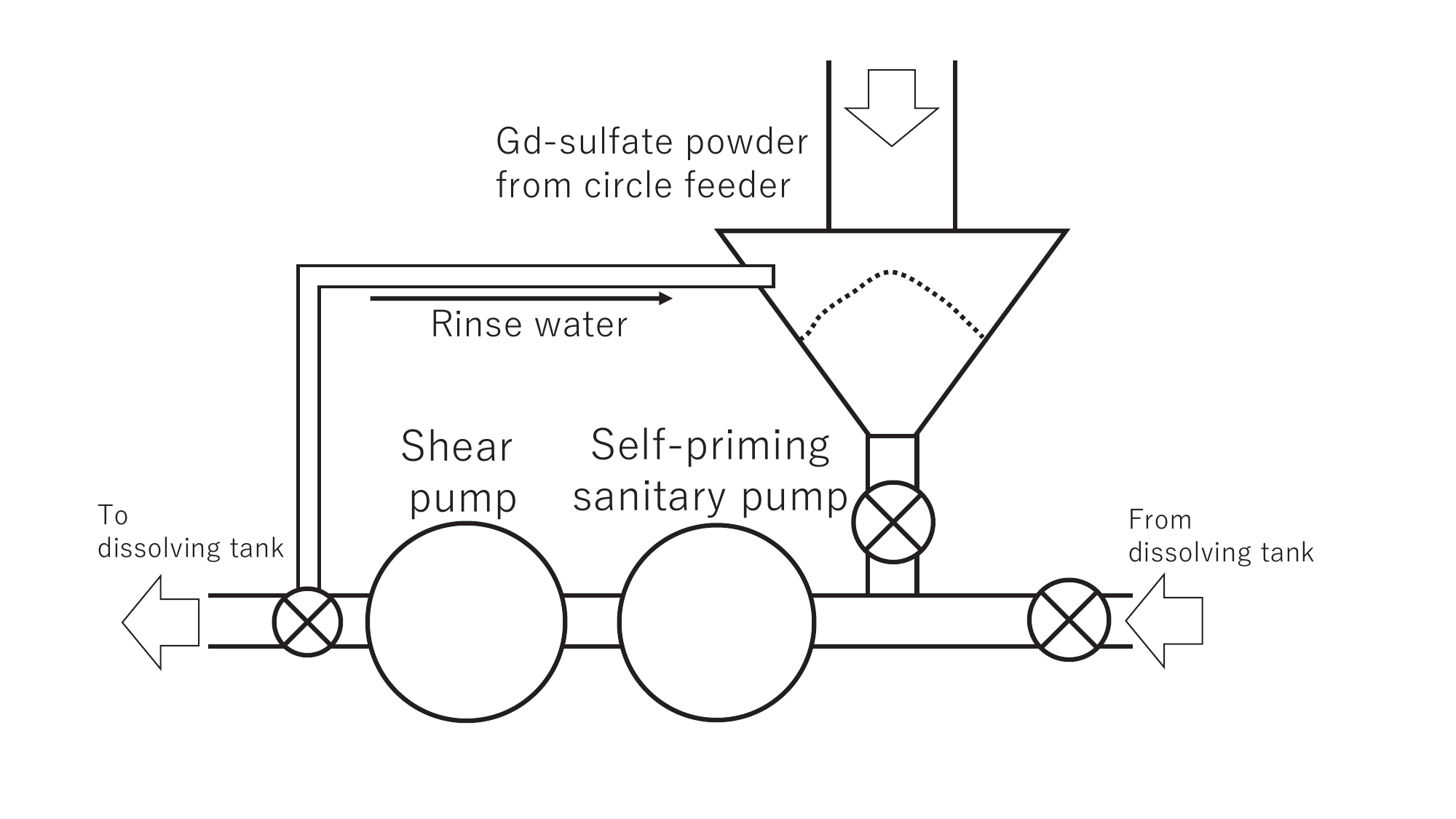}
\caption{Schematic diagram of the shear blender.}
\label{fig:dissolving}
\end{figure}

The Gd-dissolving system is divided into two parts, responsible for the transportation of the \GdSOw powder to a weighing hopper and then dissolving it into water in a buffer tank (the dissolving tank) via a circle feeder (Fig.~\ref{fig:gd-water-sys}). 
This setup allows for measured amounts of the Gd compound to be dissolved in order to achieve the desired concentration prior to returning the solution to SK.
During dissolving, a fraction of the SK water being continuously recirculated is fed into a buffer tank (the solvent tank) before it is transferred to the dissolving tank to receive \GdSOw from the shear blender.
Water from the dissolving tank is used to mix and deliver the \GdSOw powder in the shear blender before returning to the dissolving tank (Fig.~\ref{fig:dissolving}).
It is the circulation of water between the blender and dissolving tank that mixes and dissolves the powder into the SK water. 
The resulting solution is sent to the solution tank before being pumped to the pre-treatment system.



\subsection{Pretreatment System}\label{sec_pretreatment}
The pretreatment system and the main recirculation system were designed for SK-Gd in order to remove substances other than gadolinium and sulfate ions from the water. 
This design was tested with a 200-ton demonstrator experiment, the EGADS detector~\cite{egads:2020}.
Following a passive one micron filter, in the first active stage of pretreatment  gadolinium-loaded water from the solution tank is irradiated with ultraviolet (UV) light of sufficient energy to oxidize carbon and other compounds via a UV total organic carbon reduction lamp (TOC lamp).
Any ionized impurities, including some radioactive impurities such as uranium and radium, are then removed by downstream ion-exchange resins.

To remove positively charged impurities, and radium ions in particular,  a strongly acidic cation exchange resin, AMBERJET\texttrademark 1020 \cite{organo} (``C-Ex Resin'' in Fig.~\ref{fig:gd-water-sys}) is used.
For the SK-Gd water system, this resin has been modified to contain gadolinium as the ion exchange group such that the resin's cation exchange action never results in a loss of dissolved Gd content;
\GdSOw powder of the same radiopurity as the primary solute was used to generate this custom resin. 

In addition, negatively charged impurities are removed using a strongly basic anion exchange resin AMBERJET\texttrademark 4400 \cite{organo2} (``A-Ex Resin'' in the figure).
This resin was introduced in order to remove uranium, which forms a negatively-charge uranyl complex in solution.
As with the cation resin, this anion resin has been specially prepared, in this case with sulfate ions as the ion exchange group to prevent loss of sulfate from the Gd compound; electronic grade  H$_2$SO$_4$ was used to produce the resin.
In the pretreatment system the ratio of cation and anion exchange resin volumes is $1:1$.

The Gd-loaded water is then sent to a UV sterilizer and a series of filters to remove bacteria introduced during the dissolution process. 
It should be noted that the pretreatment system is used only during  dissolution and is otherwise bypassed during normal recirculation operation.


\subsection{Water Recirculation System}
During both gadolinium loading and subsequent data-taking, the Gd-loaded water in SK is continuously recirculated and purified to maintain high transparency. 
As shown in Fig.~\ref{fig:gd-water-sys}, there are two copies of most elements of the recirculation system installed in parallel such that water can be circulated at a  flow rate of 60~m$^3$/h in either line or 120~m$^3$/h when both are operated simultaneously. It is therefore possible to maintain continuous recirculation with one line while performing maintenance on elements of the other.
The configuration of the recirculation system is similar to that of the pretreatment system, but with the 0.2~micron filters replaced by ultrafiltration (UF) modules as the final filtration stage.
Additionally, the ratio of cation to anion exchange resin volumes is $1:2$.
In order to maintain precise control of the water temperature, heat-exchange units (HE) are installed immediately after the TOC lamps, just after the filtration stage, and just before the
water is returned to the detector.
Immediately before this final HE, a membrane degasifier (MD) is installed in order to remove Rn dissolved in the water.

Note that water rejected by the UF modules is returned to the first buffer tank in Fig.~\ref{fig:gd-water-sys} for reprocessing so that no water is discarded by the system. 
Loss from the system is therefore limited to evaporation from the SK tank itself, which is less than a few liters per day, and from the membrane degasifier at a rate of between 30 and 60~$\ell$/day depending on the flow rate. 
Water can be re-supplied from sources within the Kamioka mine as needed.
The system achieves stable circulation at a fixed flow rate from the return pumps by automated control of the water levels in the two buffer tanks in the figure. 


\section{Water Flow in the Tank} 
\label{sec:water_flow}
\begin{figure}[htb]
\centering\includegraphics[width=0.98\linewidth]{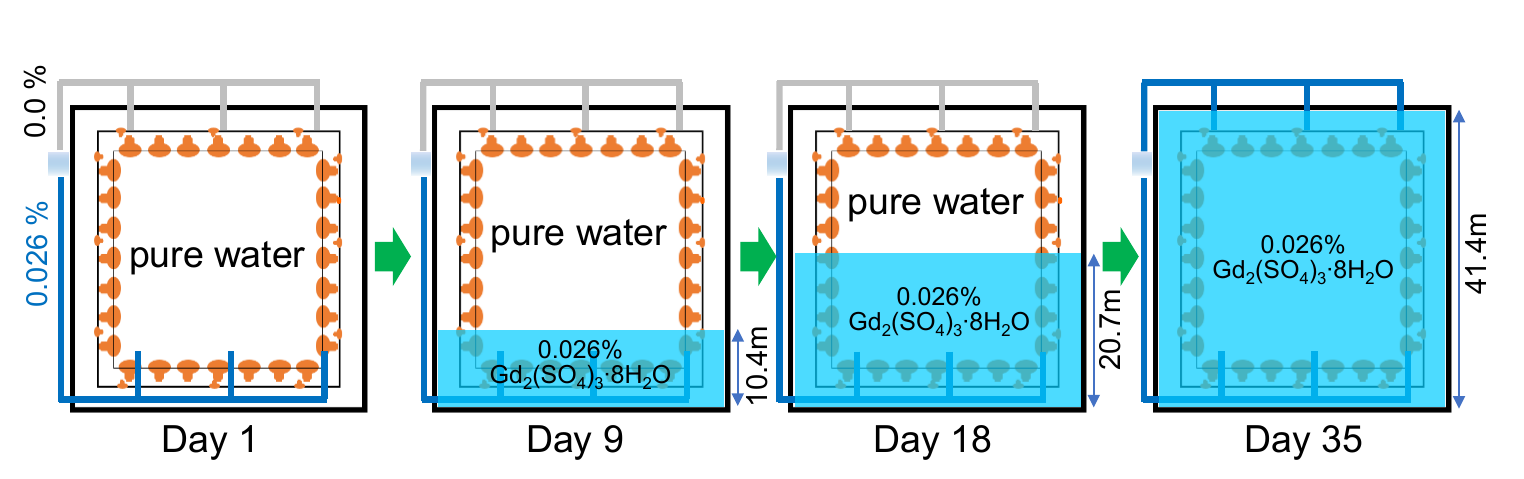}
\caption{Gadolinium loading scheme.}
\label{fig:gd_loading_scheme}
\end{figure}

The \GdSOw powder was dissolved into SK while recirculating water from the SK tank through the water system described in the previous section. 
In order to efficiently achieve a uniform concentration of Gd in the tank, we injected Gd-loaded water at the bottom of the tank while removing pure water from its top as illustrated in Fig.~\ref{fig:gd_loading_scheme}.
This was achieved by precise control of the supply water temperature and adjustment of the water flow at various water inlets and outlets on the tank. 
A brief description of the water flow control for the Gd loading is given in this section, whereas a more general description of the water flow in the  Super-Kamiokande detector can be found in \ref{Append:2}.

\subsection{Temperature Control}

The temperature of the water being sent from the water system to the tank (known as ``supply water'') is controlled at a precision better than 0.01~$^\circ$C.  This is accomplished by using a quartz thermometer which provides feedback to the heat-exchange unit installed in the final section of pipe before water is returned to the tank~\cite{Sekiya:2015rla}.
Prior to Gd loading, the water temperature in the tank was raised by setting the supply water temperature to 13.90~$^\circ$C at Temperature Control Unit B in Fig.~\ref{fig:gd-water-sys}. 
Water was recirculated under these conditions for about 45 days.
Afterwards the supply temperature was lowered to 13.55~$^\circ$C to begin the Gd loading. 
This created an additional density difference between the  Gd-loaded water and the pure water in the tank beyond that caused by the compound itself.
In this way the spatial profile of the Gd-loaded water could be monitored by measuring the water temperature at different positions in the detector.

\subsection{Water Injection and Extraction}

\begin{figure}[htb]
\centering\includegraphics[width=0.85\linewidth]{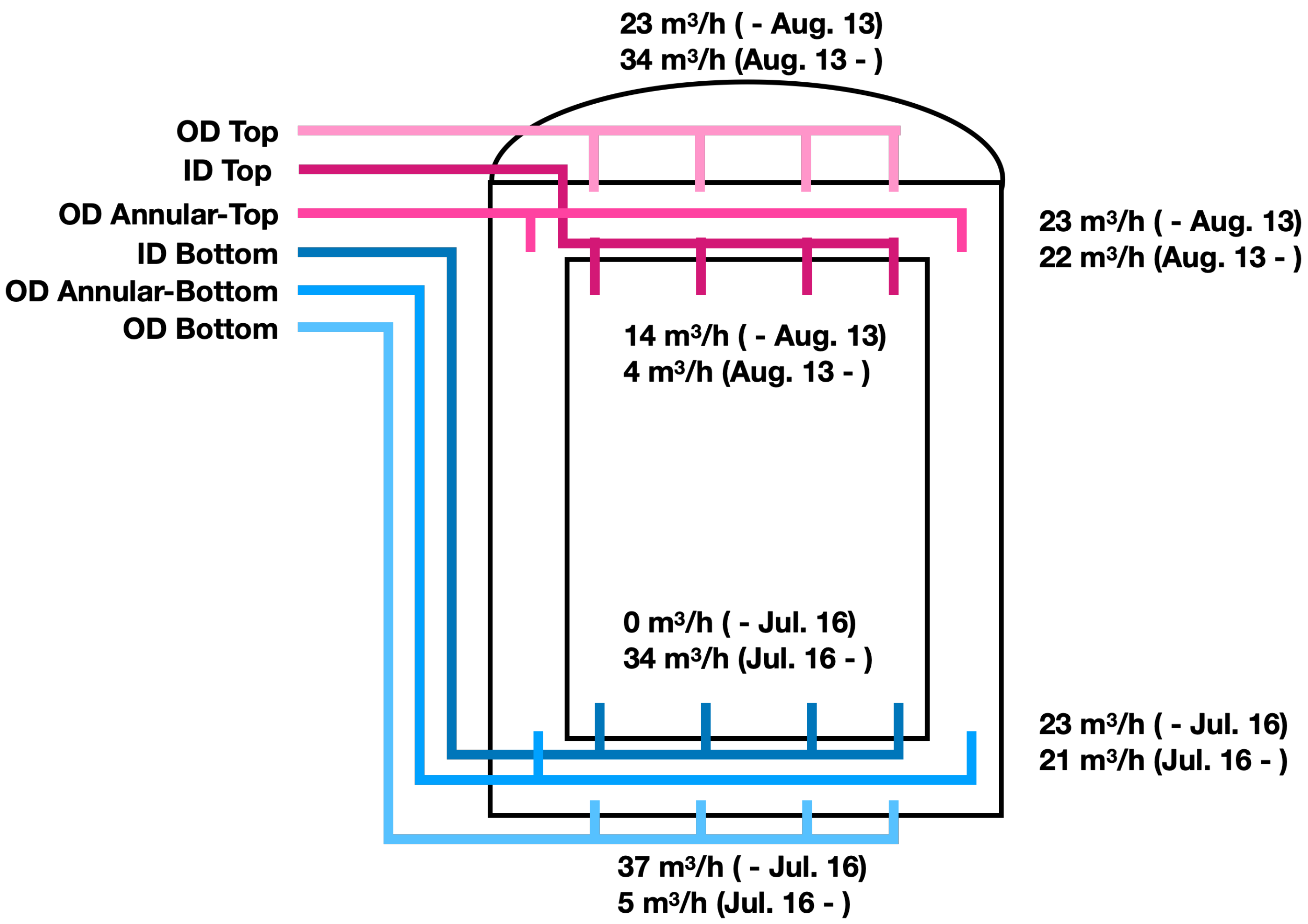}
\caption{Water flow pattern in the SK tank during the Gd dissolving in 2020.}
\label{fig:sk_water_flow_dissolving}
\end{figure}

Fig.~\ref{fig:sk_water_flow_dissolving} shows a schematic of the water piping and approximate flow rate at each location in the inner detector (ID) and outer detector (OD) regions of the tank. 
This piping was newly installed during a major in-tank refurbishment of the SK detector in 2018 and 2019 conducted in preparation for Gd loading. There are 12 inlets at both the top and bottom parts of the ID, eight in the annulus of the OD near the top and bottom, and four in the top and bottom OD endcaps. 
The end of each pipe is made of 50A polyvinyl chloride (PVC) tubing.
In order to suppress convection due to the vertical flow of injected vertical water at the bottom, diffuser caps made of 14~cm diameter stainless-steel plates were installed in February 2020 on each of the bottom ID and bottom annular OD pipe outlets (see \ref{Append:2}).  Diffuser cap installation was accomplished while the detector was filled with water through the use of a remotely-controlled submersible vehicle with a robotic arm~\cite{mes}.

In general, water was injected near the bottom and extracted near the top of the tank. 
In order to keep the height of Gd-loaded water uniform across the entire detector region,
the ratio of water flow in the combined ID and OD bottom region relative to that of the OD annular region was set to be roughly equal to the ratio of the ID and OD cross sections.
During the initial stage of Gd loading, most of the  water was injected through the OD bottom and OD annular outlets. After the rising front of Gd-loaded water reached the level of the bottom ID PMTs (on July 16, 2020), the water flow pattern was changed so that most of it was injected through the ID bottom and OD annular outlets.
A similar adjustment was also made for outlets at the top of the OD and ID when the Gd-loaded water reached the top ID PMTs (August 13, 2020).

\section{The Gd-loading}
\label{S:4}
\subsection{Gd Powder Specification}
\label{S:4-1}
The \GdSOw for SK-Gd had to fulfill stringent requirements regarding cleanliness to ensure it was free from problematic levels of contamination. 
As shown in Table~\ref{tab:RIinGd}, criteria for radioactive impurities were set so that the additional event rate due to radioactive impurities in the powder -- even after loading up to the final goal of 0.1\% Gd concentration -- would be less than the unloaded background rates for SK's solar neutrino measurements or supernova relic neutrino (SRN) searches~\cite{pablo:2017phD}.  In other words, less than a doubling of the pure water background rates for these analyses was deemed acceptable.
In order to meet the requirements, chemical processing procedures were developed by an extensive R\&D program~\cite{egads:2020,Ito:2020ptep}.
The calculations underlying the specifications of this ultraclean \GdSOw powder are described elsewhere~\cite{pablo:2017phD,Ito:2020ptep}.

\begin{table}[htb]
	\centering
	\caption{Criteria of radioactive impurities in \GdSOw powder.}
	\label{tab:RIinGd}
    \vspace*{0.1cm}
    \begin{tabular}{cccc}
	    \hline 
		Chain & Isotope & Criterion [mBq/kg] & Physics target \\
		\hline
		\hline
		\multirow{2}{*}{$^{238}$U}  & $^{238}$U  &  $<$ 5  & SRN  \\
 			                    	& $^{226}$Ra   & $<$ 0.5 & Solar \\
		\hline
		\multirow{2}{*}{$^{232}$Th} & $^{232}$Th   & $<$ 0.05 & Solar \\
			                     	& $^{228}$Ra   & $<$ 0.05 & Solar \\
		\hline
		\multirow{2}{*}{$^{235}$U} & $^{235}$U     & $<$ 30 & Solar \\
			                     	& $^{227}$Ac/$^{227}$Th & $<$ 30 & Solar \\
		\hline
	\end{tabular}
\end{table}

The final production was conducted in batches of 500~kg. Every 500~kg lot of \GdSOw powder was delivered in a special EVOH (ethylene-vinylalcohol copolymer) lined flexible container. Each batch was carefully screened at one (or more) of our collaborating underground scientific laboratories: Boulby in the UK, Canfranc in Spain, and the Kamioka Observatory in Japan~\cite{ikedaLRT}.  


Since the powder contained an average of 2.5\% additional water left over from processing, the 13.2~tons of dissolved powder implies that the mass of \GdSOw itself is 12.9~tons.
This amount should yield a gadolinium (Gd) concentration in the SK tank of 0.011\%, equivalent to an anhydrous gadolinium sulfate ($\rm Gd_2(\rm SO_4)_3$) concentration of 0.021\%. 



\subsection{Loading Scheme}
During \GdSOw powder loading, the SK return water flow (after the ``Return pumps'' in Fig.~\ref{fig:gd-water-sys} ) was set to 60~m$^3$/h.
Of that, 48~m$^3$/h was directed to the main return line, while the rest was divided between the dissolving system (1.33~m$^3$/h) and its bypass line (10.67~m$^3$/h).
In the dissolving system, the following powder dissolution processes were repeated:
\begin{enumerate}
  \item Transfer 750~$\ell$ of the solvent -- the return water from SK -- from the solvent tank to the dissolving tank.
  \item Using the circle feeder, put 8.2~kg of powder from the weighing hopper into the shear blender.
  \item Circulate the solution between the blender and the dissolving tank for 15~minutes.
  \item Stop the circulation, and transfer the solution from the dissolving tank to the solution tank. After that, return to process~\#1.
\end{enumerate}
Each one of these cycles took about 30 minutes to complete, and the cycle ran continuously during the Gd loading period except when powder was being supplied to the weighing hopper.
During this powder supplying process, the dissolution process itself was briefly interrupted, but the Gd loading of SK from the solution tank continued unabated.
The dissolving cycle initially produced 0.9\% \GdSO water, which was then merged with the water from the bypass line to become 0.1\% \GdSO. This in turn passed through the pre-treatment system (see Section~\ref{sec_pretreatment}) and was 
finally merged with the water from the main return line to produce the  0.02\% \GdSO water sent to SK.

\subsection{Gd Loading History}
\label{S:4-3}
\begin{figure}[htb]
\centering\includegraphics[width=0.8\linewidth]{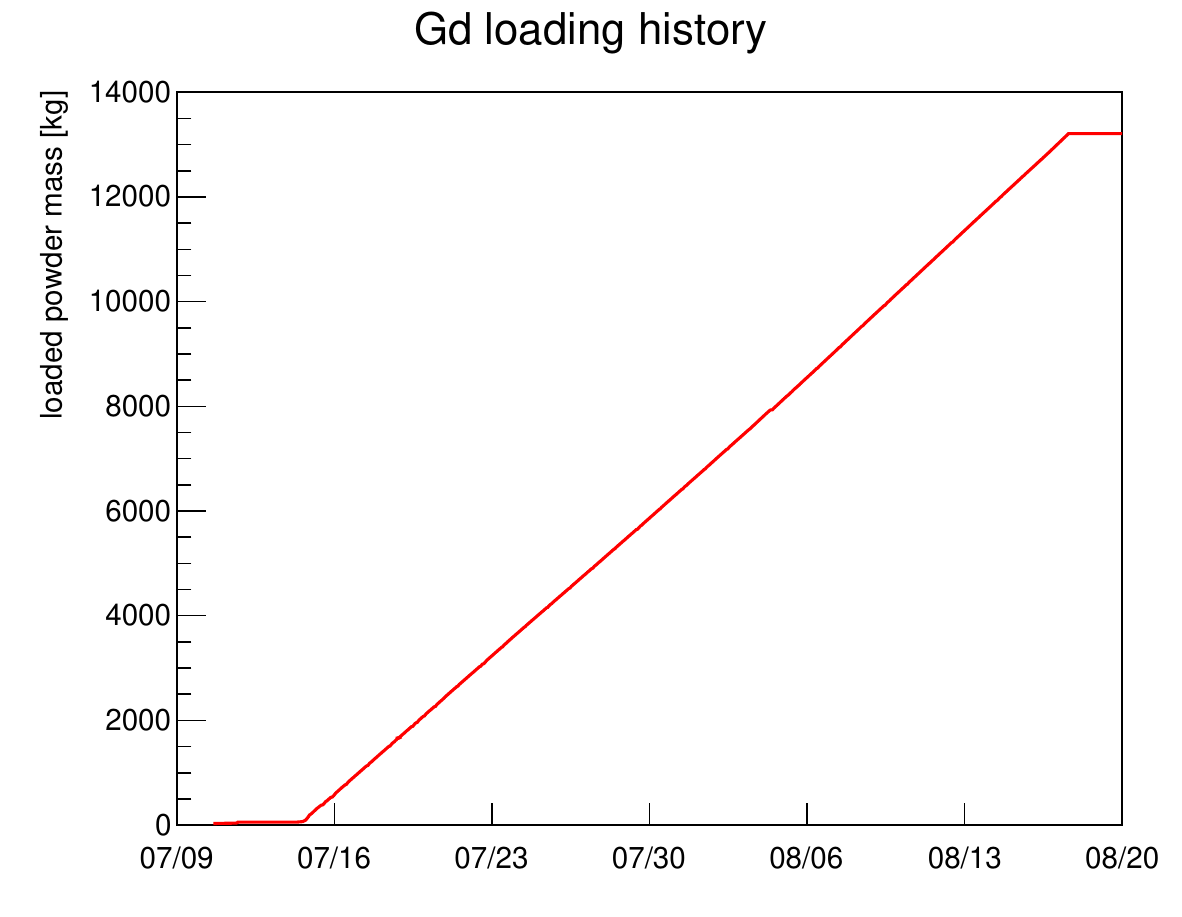}
\caption{The history \GdSOw powder loading in SK during 2020. On August 17, the amount of dissolved \GdSOw powder reached the target of 13.2~tons.}
\label{fig:gd_loading_history}
\end{figure}

Fig.~\ref{fig:gd_loading_history} shows the cumulative mass of loaded \GdSOw powder for each cycle.
It can be seen that the loading was very stable up to the target value of 13.2~tons, which corresponded to 5426~kg of Gd in the 49.47 kilotons of water in the SK tank and the water system. We assume 0.5\% errors on both the amount of Gd and water; the estimated Gd concentration in the tank should therefore be 109.7$\pm$0.7 ppm (we refer to this Gd concentration as ``0.011\%'' or ``0.021\%'' \GdSO concentration.).

The amount of \GdSOw powder added in one cycle was changed from 8.2~kg to 8.7~kg on August 4, 2020, and 
Gd loading was completed on August 17, 2020.
Then the water recirculation at the rate of 120~$\rm m^3/h$ was started on August 18, 2020.


\subsection{Water Transparency} 
As mentioned in Section \ref{S:2}, one of the main purposes of the water system is to keep the SK water transparency good enough for physics analyses. Although the transparency of the Gd-loaded water itself had been demonstrated with EGADS~\cite{egads:2020}, the Gd-loading scheme, the water flow rate, and the water flow pattern in SK are different from those in EGADS. Here we report the effects of the Gd loading activity and the SK-Gd water recirculation on the water transparency. 

\begin{figure}[htb]
\centering\includegraphics[width=1.0\linewidth]{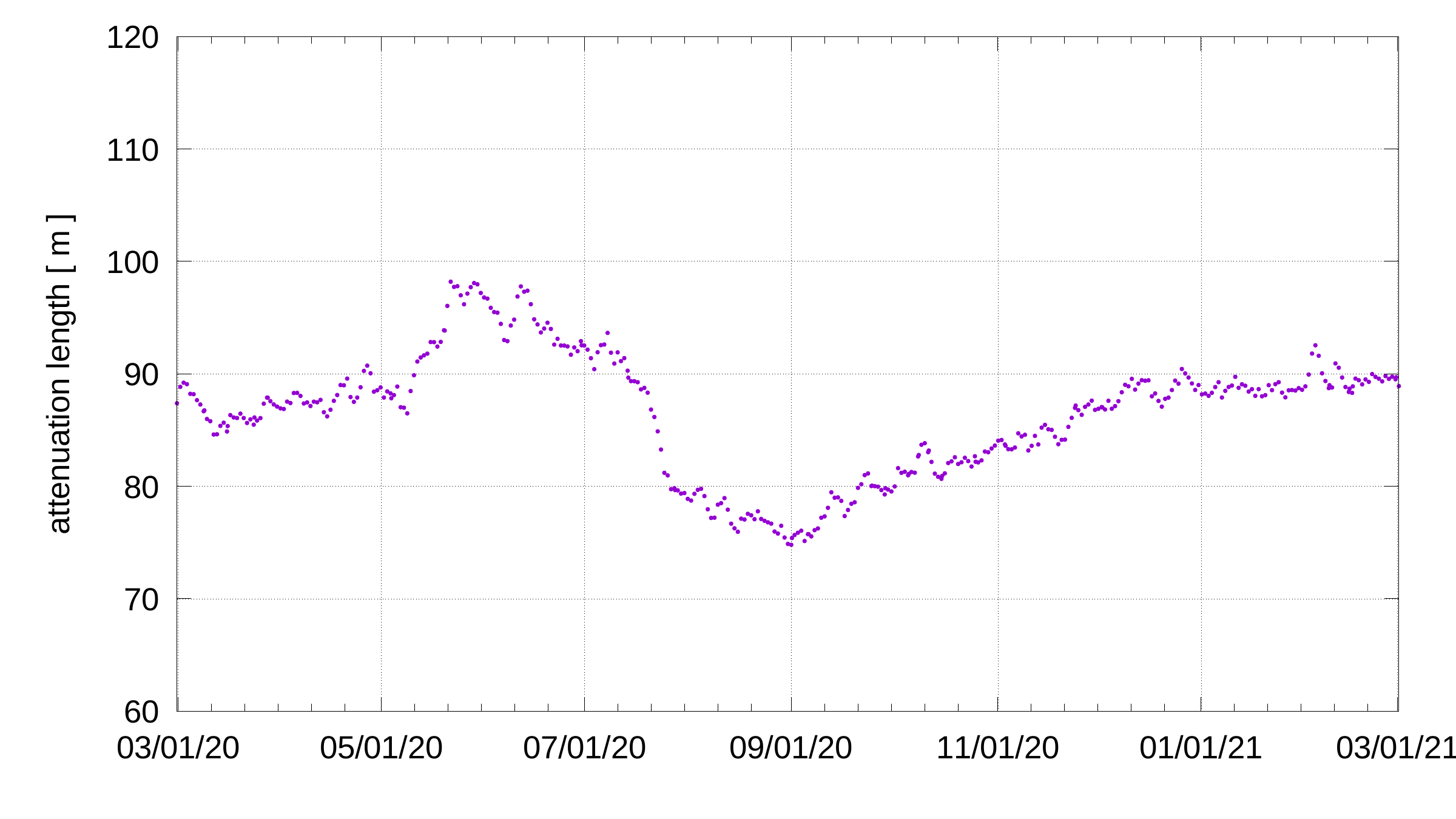}
\caption{The light attenuation length measured with cosmic ray through-going muons from March 1, 2020, to March 1, 2021. From April 29, 2020, we lowered the temperature of the supply water to the bottom of the tank in order to simulate the Gd-loading water flow described in Section~\ref{sec:water_flow} and \ref{SS:B3}. As a consequence, the water replacement efficiency and hence the transparency improved until the Gd-loading period began.}
\label{fig:trans}
\end{figure}

The time variation of Cherenkov light attenuation length in the SK water tank measured with cosmic ray through-going muons is shown in Fig.~\ref{fig:trans}. 
These results are used to correct observed PMT charges in the event reconstruction for all higher energy physics analyses such as atmospheric neutrino oscillations~\cite{Abe:2018atm}, nucleon decay searches, and long-baseline neutrino oscillations. The analysis details of the attenuation length measurement can be found elsewhere~\cite{Fukuda:2003skdet}. The attenuation length was about 90 meters just before loading Gd in the SK tank on July 14, 2020. After loading $\rm Gd_2(SO_4)_3\cdot 8H_2O$ began, the attenuation length began to decrease. Around the end of August 2020, it reached a minimum value of around 75 m. Then, it started to recover, and had returned to almost 90 m at the beginning of December, 2020.
The attenuation length became as long as it had been during the pure water phases of SK.

\subsection{Determining the \GdSOw Concentration from Direct Sampling in the SK Detector}

\begin{figure}[htb]
\centering\includegraphics[width=0.9\linewidth]{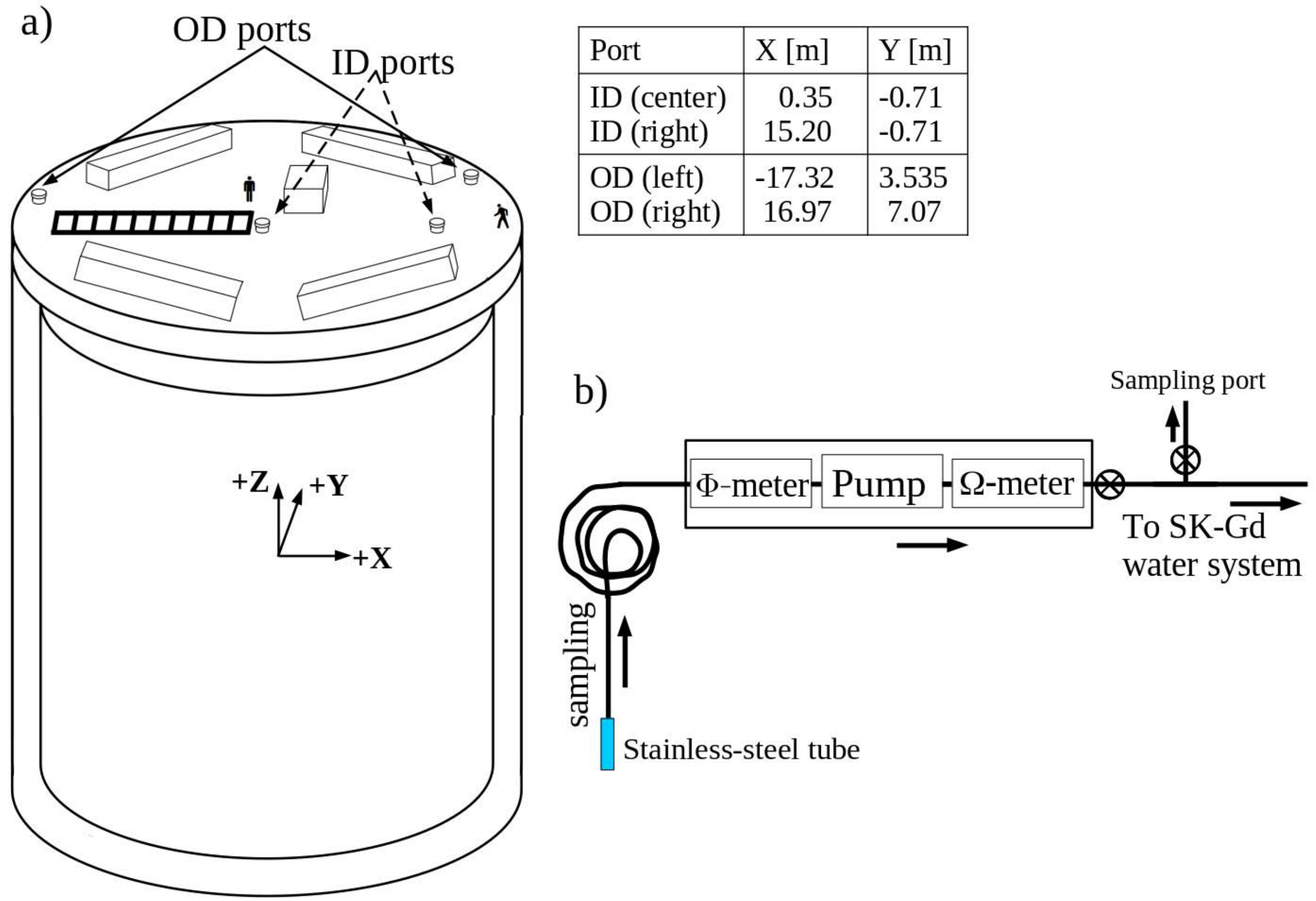}
\caption{Water sampling system: a) (left) Schematic view of the position of the sampling ports and coordinate system in SK. b) (right) Schematic view of the sampling system.
$\Phi$-meter is a flow meter and $\Omega$-meter is a resistivity meter.}
\label{fig:sampling_ports_and_system}
\end{figure}

During the Gd loading period, daily water samples were taken from calibration ports as depicted in  Fig.~\ref{fig:sampling_ports_and_system}. In the ID, two positions were used: the center-most port and the closest port to the OD in the +X axis. In the OD, two positions were also employed: one in the +X and another one in the $-$X hemisphere. 

The system to take the samples was rather straightforward (see Fig.~\ref{fig:sampling_ports_and_system}). The sampling probe consisted of a flexible plastic tube 40 m long in which the last 25 cm were replaced by a stainless steel tube. This metal provided sufficient additional weight to compensate for the mild buoyancy of the flexible tube, and thereby kept the sampling probe straight even when it was completely submerged in the SK detector. 
The sampling probe was connected to a sampling system that consisted of three devices: a flow meter, a pump, and a conductivity meter which also functioned as a thermometer. At the end of this system there was a sampling port from which samples could be collected for later measurement of the \GdSOw concentration with an Atomic Absorption Spectrometer (AAS). Excess water -- in particular that used for flushing the sampling tube at each new position -- was sent back into the SK water system, cleaned, and returned to the detector. Because a complete sampling operation (extracting water from different depths at a given calibration port) took about four hours, several measures were employed to avoid having to disturb the usual data taking of SK. Most of these measures focused on avoiding light leaks into the detector.

The general procedure to take samples was as follows: 
\begin{enumerate}
\item The sampling probe was placed inside a darkened black tent above the chosen sampling port.
\item The probe was then immersed about 2 m into the SK detector and the pump was turned on to make sure that there was no air trapped in the system.
\item After waiting to allow the sampled water from this position to completely flush the system, the conductivity would become constant.
\item Water flow, conductivity, and temperature were then recorded. A water sample was taken in a small (about 15 m$\ell$) bottle, upon which the position in the detector (sampling port and depth) and date were written.
\item The sampling tube was lowered 2 m further down into the detector.
\item Back to Step 3 until the lowest depth in the given port (Z=$-16$ m in the ID and Z=$-18$ m in the OD) was reached and sampled. The sampling tube was then pulled out of the detector and the system shut down.
\end{enumerate}


Before loading Gd, the SK water conductivity was close to 55 nS/cm, the theoretical minimum, but after the addition of gadolinium and sulfate ions the conductivity rose to about 168 $\mu$S/cm. Because the pretreatment and water recirculation systems were specifically designed to remove all impurities except for these two ions, this should have made the conductivity a good indicator of the \GdSO concentration.

\begin{figure}[htb]
\centering\includegraphics[width=0.9\linewidth]{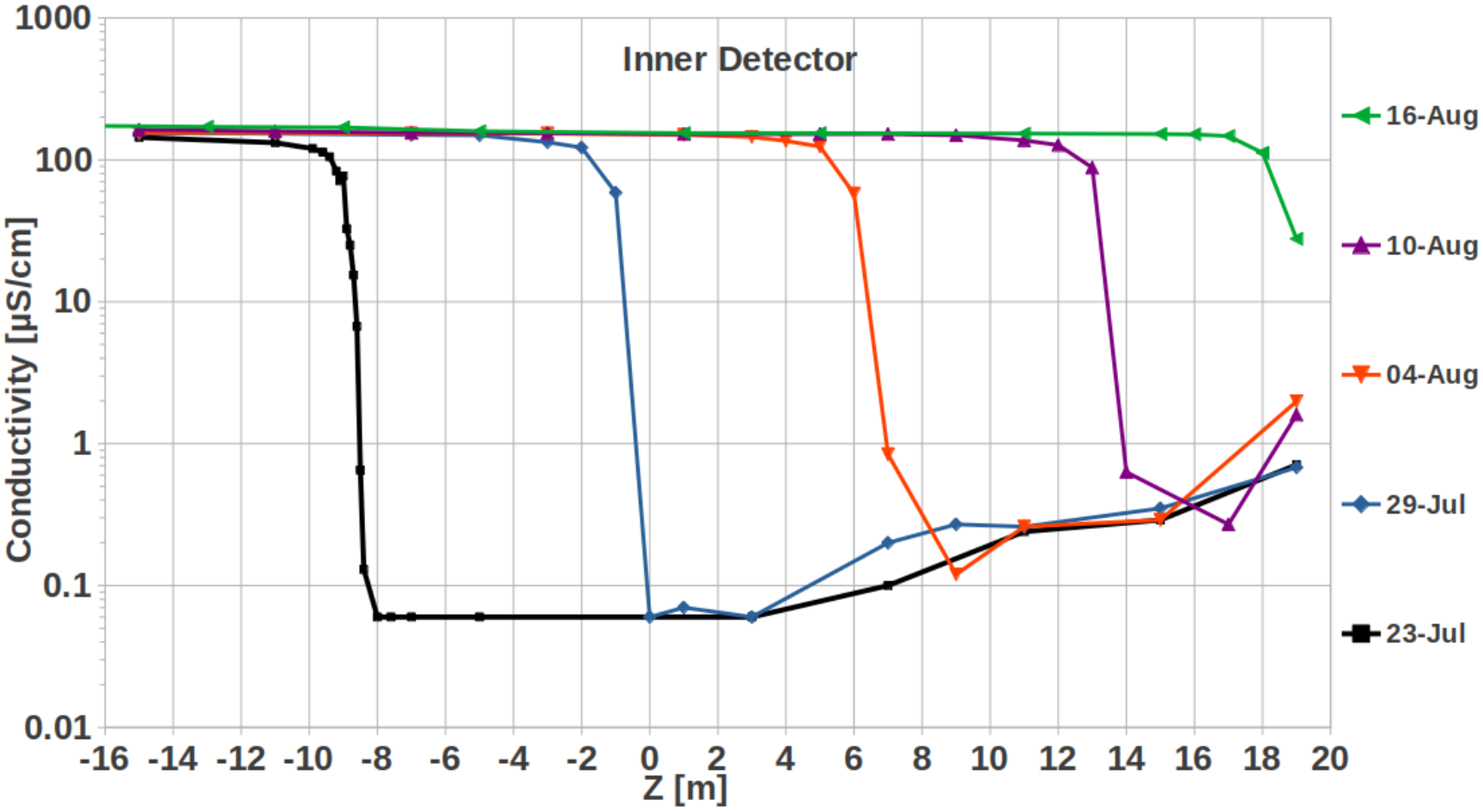}
\caption{Conductivity vs. Z position in the ID during Gd loading. Z=+19 m is inside the calibration guide pipe.}
\label{fig:Conductivity-ID-loading}
\end{figure}

Figure~\ref{fig:Conductivity-ID-loading} shows the time evolution of conductivity in the ID center port position as a function of Z during the 2020 loading of Gd into the detector; the other ID port yielded very similar results and so is not shown. The gap between July 21 and July 29 was due to a problem in the sampling system. Gd loading started on July 14, but no significant change in the conductivity was observed until July 19. We can see a rather sharp boundary about 2 m thick existed between the Gd loaded layer ($>$100 $\mu$S/cm) and the pure water layer ($<$1 $\mu$S/cm). In agreement with the rate of water flow in the detector, this boundary advanced about 1.5 m/day. 

\begin{figure}[htb]
\centering\includegraphics[width=0.9\linewidth]{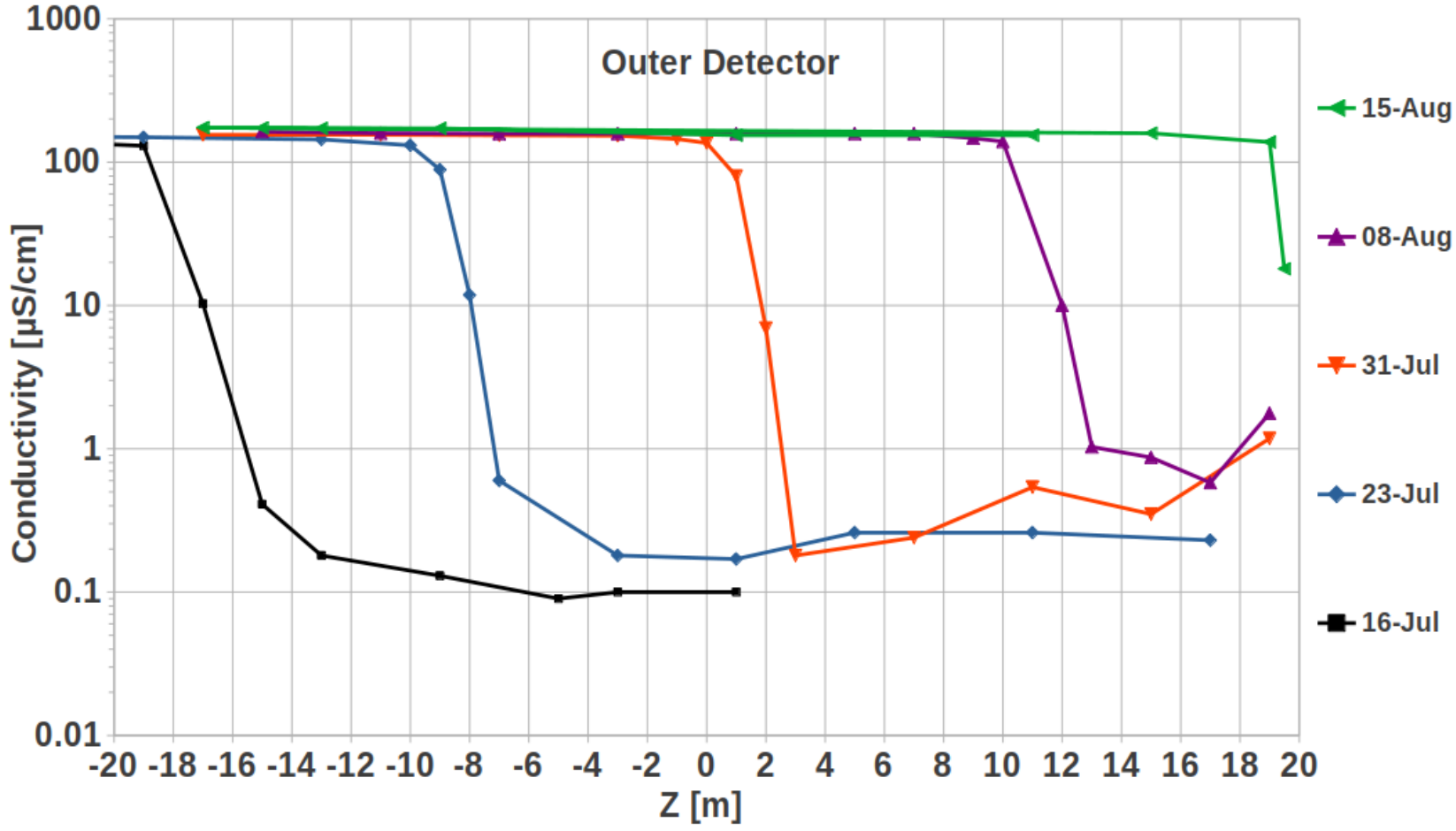}
\caption{Conductivity vs. Z position in the OD during Gd loading. Z=+19 m is inside the calibration guide pipe.}
\label{fig:Conductivity-OD-loading}
\end{figure}

Figure~\ref{fig:Conductivity-OD-loading} shows the conductivity in the OD $-$X port; as was the case in the ID, the other OD port yielded similar results and so is not shown. The general behavior (boundary thickness and speed) was the same as in the ID, but we can observe that the boundary was about 1 m higher than in the ID due to the higher water flow in the OD. It can be seen that just after Gd loading was completed the conductivity, i.e. the concentration, was not homogeneous yet.

As observed above, conductivity should have been a good indicator of the \GdSOw concentration. However, it was not a direct measurement; if anything unexpected had occurred and other ions were present in large amounts, then the conductivity would not have been a good indicator of the \GdSOw concentration. 

To make sure the Gd concentration was indeed as expected, most water samples were analysed with an AAS. A Hitachi polarized Zeeman AAS of the ZA3000 series was used with Pyro Tube HR cuvettes. The samples were atomized at 2700 $^\circ$C and illuminated with a Gd hollow-cathode lamp. For the calibration standard, samples containing 10 and 20 ppm \GdSOw were carefully prepared using the same \GdSOw powder that was used to load the SK detector. Calibration was performed before every series of measurements and also in between them to monitor the measurement stability. The uncertainty of a single measurement was about 3.5$\%$

\begin{figure}[htb]
    \centering\includegraphics[width=0.9\linewidth]{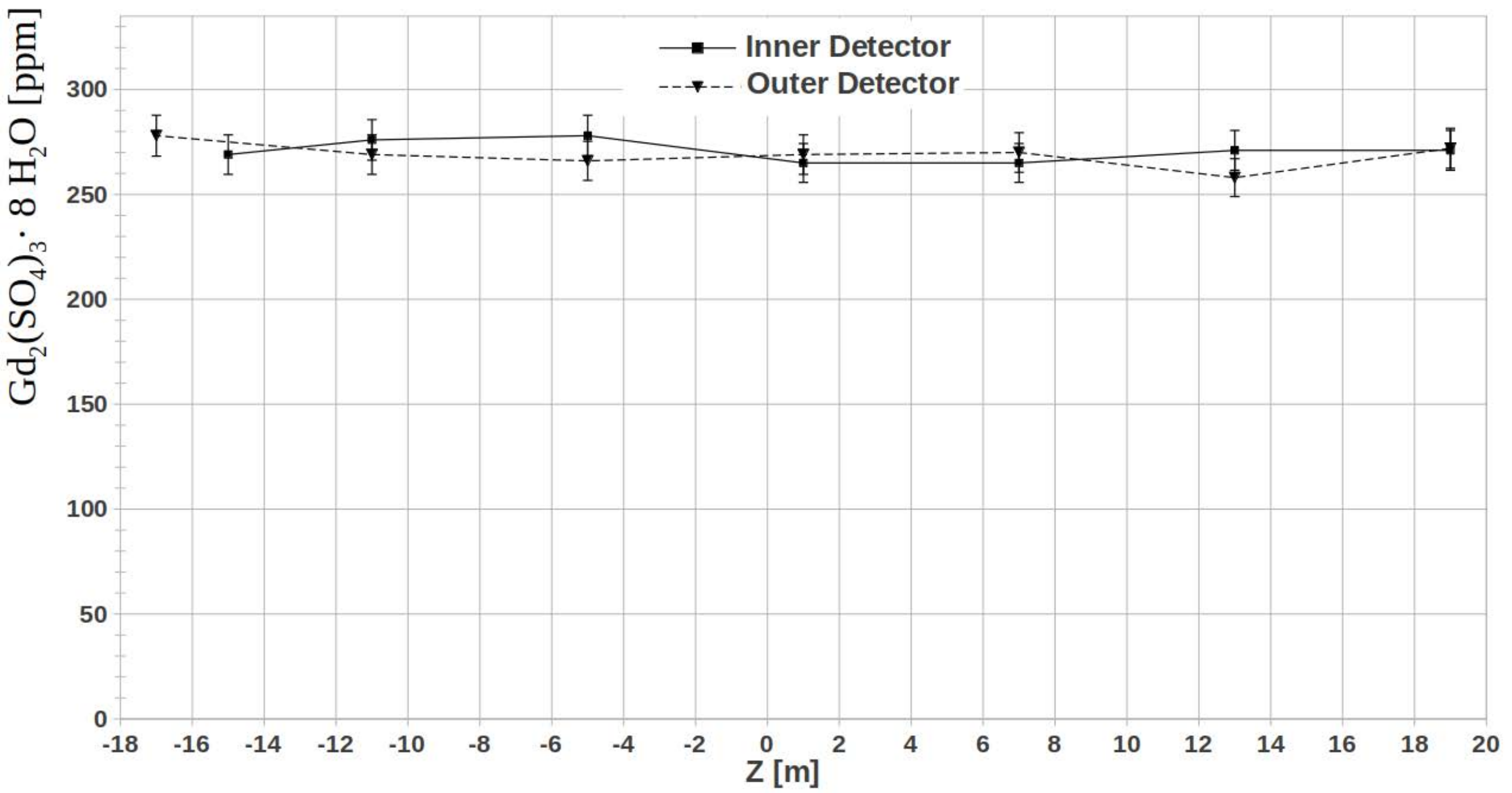}
\caption{Latest AAS-measured \GdSOw concentration vs. Z position in the ID and OD from samples taken on March 25, 2021, showing the complete homogenization within the detector.}
\label{fig:Concentration-IDOD-last}
\end{figure}

AAS measurements were conducted on samples taken during and after the 2020 period of Gd loading. 
This allowed us to to confirm that the concentration was becoming homogeneous using  localized conductivity measurements as well as the concentration final value.
Figure~\ref{fig:Concentration-IDOD-last} depicts the latest available AAS measurements (from samples taken on March 25, 2021) in the ID and OD regions, showing that the concentration had become completely homogeneous with an average \GdSOw concentration value of 271 $\pm$ 4 ppm. Taking into account the non-Gd components in the \GdSOw powder, this AAS result translates to an entirely uniform Gd concentration in the SK water of 114 $\pm$ 2 ppm.
This value is slightly larger than the estimation from the
weight described in Section \ref{S:4-3}, but the difference is within the errors of the AAS measurement and the final concentration estimation.


\subsection{Evaluation of Gadolinium Concentration by Am/Be Neutron Source}
The thermal neutron capture cross section of gadolinium is about $10^5$ times larger than that of hydrogen as shown in Table~\ref{tab:GdCrossSection}.
Thus, an increase of the Gd concentration in water results in a significantly shorter neutron capture time constant even at small concentrations in mass fraction.

\begin{table}[htb]
	\centering
	\caption{The abundance of isotopes in natural gadolinium and their thermal neutron (25~meV) capture cross sections for the Gd(n,$\gamma$)Gd reaction. Thermal capture cross sections and natural abundance of hydrogen (proton), oxygen, and sulfur -- the only other elements present in significant quantities in SK's  \GdSO-loaded water -- are also shown~\cite{IUPAC2016, ENDF7}.
	}
	\label{tab:GdCrossSection}
    \vspace*{0.1cm}
	\begin{tabular}{rrrr}
		\hline
		Isotope
		& \shortstack{Natural abundance\\ratio [\%]}
		& \shortstack{Thermal capture\\cross section [barn]}
		\tabularnewline
		\hline
		${}^{152}$Gd & 0.20  & 740 \tabularnewline
		${}^{154}$Gd & 2.18  & 85.8 \tabularnewline
		${}^{155}$Gd & 14.80 & 61100 \tabularnewline
		${}^{156}$Gd & 20.47 & 1.81 \tabularnewline
		${}^{157}$Gd & 15.65 & 254000 \tabularnewline
		${}^{158}$Gd & 24.84 & 2.22 \tabularnewline
		${}^{160}$Gd & 21.86 & 1.42 \tabularnewline
		\hline
	    ${}^{1}$H & 99.99 & 0.33 \tabularnewline
		${}^{16}$O & 99.76 & 0.0002 \tabularnewline
		${}^{32}$S & 94.85 & 0.53 \tabularnewline
		\hline
	\end{tabular}
\end{table}

\begin{figure}[htb]
	\centering\includegraphics[width=0.8\linewidth]{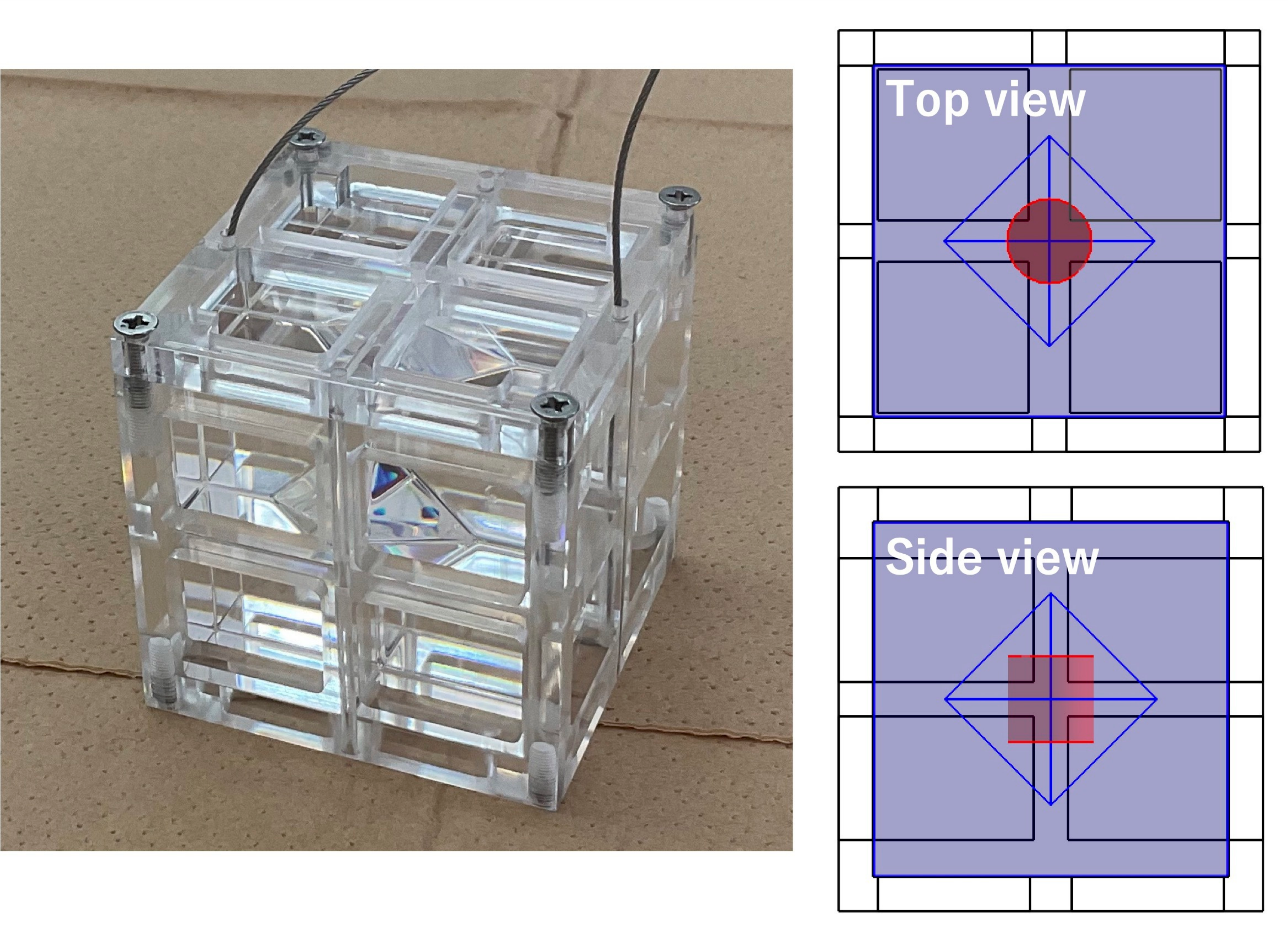}
	\caption{
	A picture of the Am/Be + BGO source (left) and its implementation in a Geant4 simulation (right).
	The Am/Be source is located at the center of eight BGO crystals, each one of which is 2.5$\times$2.5$\times$2.5~cm. A corner of each crystal is cut out to make an inner space in which to place the Am/Be source. These crystals are themselves contained within an acrylic case.
	In the Geant4 simulation, the radioactive Am/Be is defined to be at the center of a cylindrical stainless steel container (red) which is held within the otherwise empty inner space (blue). That is surrounded in turn by the eight BGO crystals (blue-gray) and an acrylic case (white).  This assemblage  is considered to represent the overall source structure in its entirety.
	}
	\label{fig:AmBe_source}
\end{figure}

By measuring the neutron capture time constant using an Am/Be neutron source (${}^{241}\!\mathrm{Am}\rightarrow {}^{237}\mathrm{Np}+\alpha, {}^{9}\mathrm{Be}+\alpha\rightarrow {}^{12}\mathrm{C}^{*}+\mathrm{n}, {}^{12}\mathrm{C}^{*}\rightarrow{}^{12}\mathrm{C}+\gamma\mathrm{(4.4~MeV)})$ deployed in the SK detector, the concentration of gadolinium was evaluated.
The advantage of using an Am/Be source lies in the ability to identify its neutron emission due to the coincident emission of a 4.4~MeV gamma-ray.
Placing the Am/Be source within a 5$\times$5$\times$5~cm BGO crystal cube~(Fig.~\ref{fig:AmBe_source}) allowed 
the initial gamma-ray to be detected by SK from the large number of scintillation photons generated by its passage through the BGO~\cite{watanabe2009}.
Following neutron capture and subsequent emission of a cascade of gamma-rays from the excited Gd nucleus, these delayed neutron-induced events were observed as a result of the de-excitation gammas Compton scattering off electrons in the SK water and thereby producing detectable Cherenkov light. 
The total gamma-ray energy produced from neutron capture on gadolinium is typically $\sim$8~MeV, and as can be inferred from Table~\ref{tab:GdCrossSection} primarily comes from two isotopes, $^{155}$Gd and $^{157}$Gd:
	$\mbox{n}+{}^{155}\mbox{Gd}\rightarrow {}^{156}\mbox{Gd}+\gamma\mbox{'s}\ [\mbox{8.5~MeV\ in\ total}],
	\mbox{n}+{}^{157}\mbox{Gd}\rightarrow {}^{158}\mbox{Gd}+\gamma\mbox{'s}\ [\mbox{7.9~MeV\ in\ total}]$.
	
The data were taken by deploying the Am/Be source into the SK inner detector through a calibration port near the center in the X-Y plane (X=$-$3.9~m, Y=$-$0.7~m).
Three positions along the Z-coordinate were selected for periodic monitoring: Z=0~m, Z=$+$12~m and Z=$-$12~m.
When searching for neutron capture events in SK, specialized data acquisition triggers were applied:  the  so-called SHE (super-high-energy) and AFT (after) triggers~\cite{Zhang:2013tua}.
An SHE trigger was generated whenever more than 60 ID  photomultipliers tubes (PMTs) detected at least one photon within a 200~ns time window. These SHE triggers resulted in
all photons detected by PMTs during the 35~$\mu$s following the SHE trigger being recorded, while those in a subsequent 500~$\mu$s window were also recorded by a sequentially issued AFT trigger.
Gd(n,$\gamma$)Gd event candidates were then extracted from this recorded data by looking for greater than 30 active PMTs in a 200~ns time window and applying event vertex reconstruction~\cite{Abe:2016nxk}.

\begin{figure}[htb]
	\centering\includegraphics[width=0.8\linewidth]{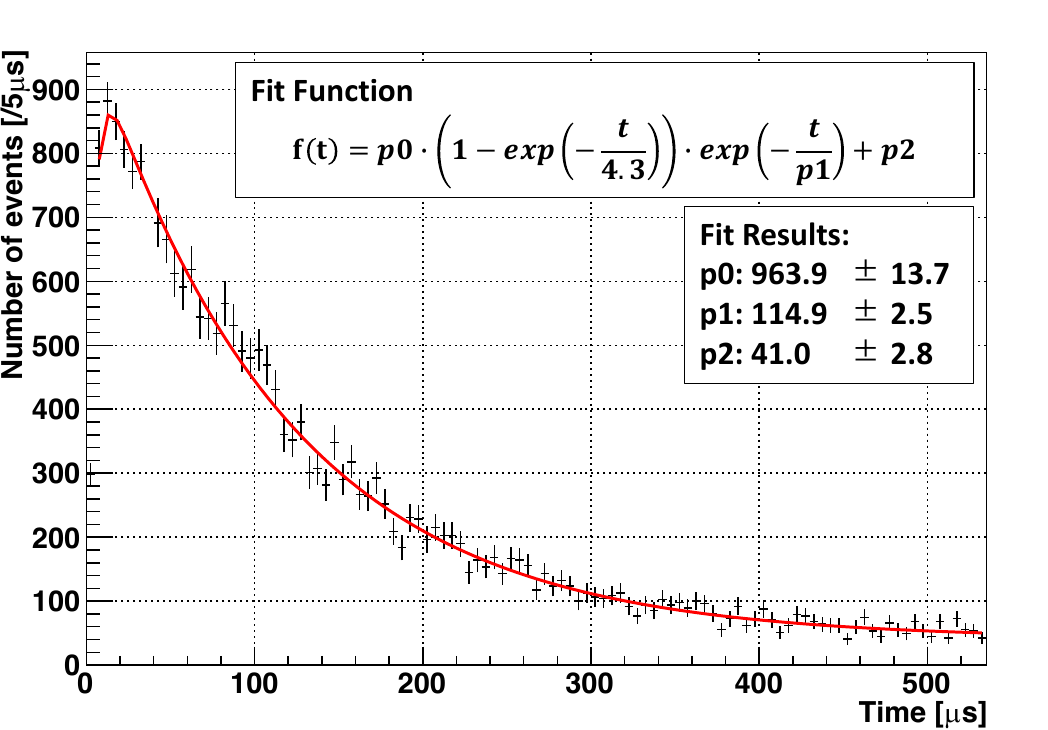}
	\caption{An example of the time distribution of neutron capture event candidates (black data points) and its fit function (red line), measured with the Am/Be source at the Z=0~m position on September 29, 2020. Each such measurement contained about 30~minutes of data with the Am/Be source deployed at various locations in the SK detector. Time zero is defined by the detection of the prompt 4.4~MeV gamma-ray BGO scintillation event. 
	The neutron capture time constant is represented by $p1$, while the thermalization time constant of 4.3~$\mu$s is derived from summed analysis of these measurements. 
	}
	\label{fig:TimeSpectrum_29thSep2020}
\end{figure}

The time distribution of neutron capture event candidates are shown in Fig.~\ref{fig:TimeSpectrum_29thSep2020}.
Event selections were applied using the following event reconstruction parameters: the 
reconstruction timing goodness $g_t$ had to be greater than 0.4, the hit pattern goodness $g_p$ smaller than 0.4, and the event vertex located within 4~m from the Am/Be source position in the SK tank. Here, the timing goodness parameter is for testing the “narrowness” of the PMT hit timing residuals. The hit pattern goodness is for testing the azimuthal symmetry of the Cherenkov cone ($g_p=0$ is perfectly symmetric, $g_p=1$ is completely asymmetric). The definitions and typical distributions of  these variables are in~\cite{Parker:sk2} [Sec. III B ]. 
In addition, the initial SHE trigger events had to include 800 to 1300 active PMTs within 1.3~$\mu$s; doing so selected 4.4~MeV gamma-ray emission from the Am/Be source.

\begin{figure}[htb]
	\centering\includegraphics[width=0.8\linewidth]{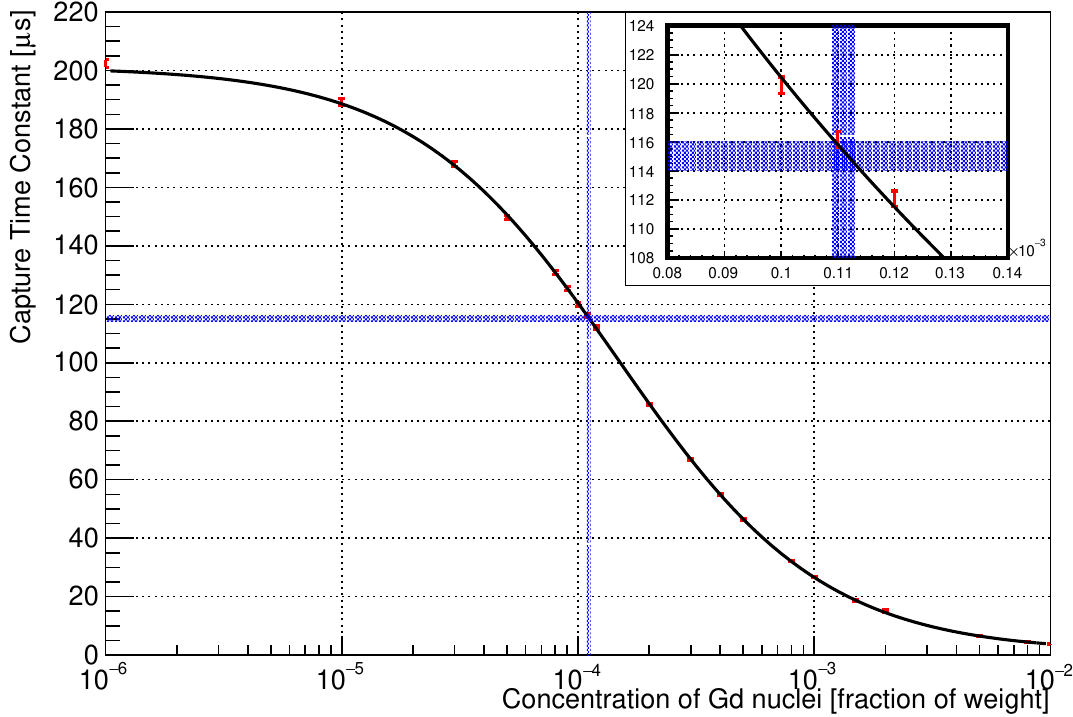}
	\caption{Neutron capture time constant as a function of the gadolinium concentration.
	The red points correspond to the Geant4 Monte Carlo simulation, while the black line corresponds to an approximate polynomial function. 
	The horizontal blue band represents the mean neutron capture time constant measured with the Am/Be source, and the vertical blue band represents the derived concentration, which is consistent with the estimation from the weight (110~ppm) described in Section \ref{S:4-3}.
	}
	\label{fig:CaptureT_Gd_MC}
\end{figure}

\begin{figure}[htb]
	\centering\includegraphics[width=1.0\linewidth]{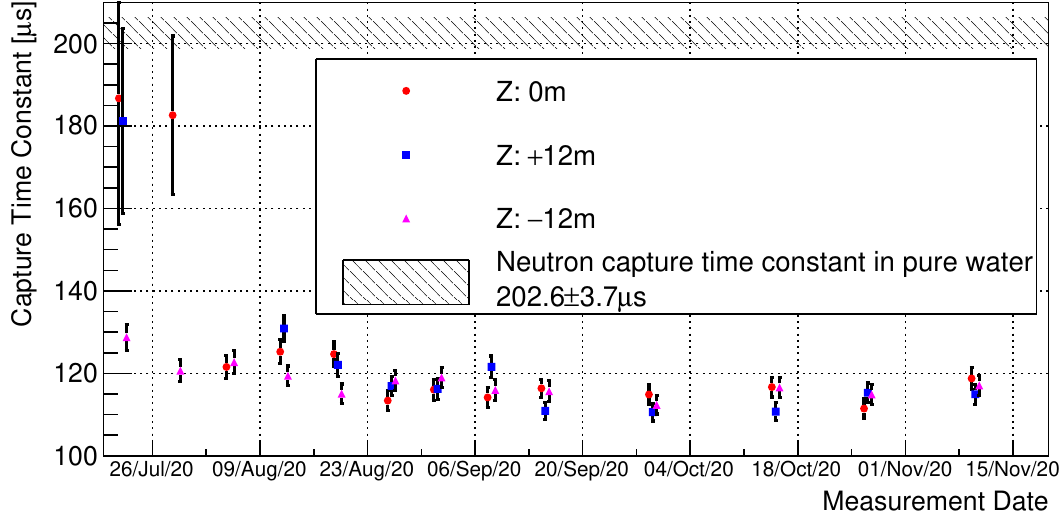}
	\caption{The history of the neutron capture time constant in SK in the latter half of 2020, obtained from the analysis of Am/Be source data.
	Three positions along the Z-coordinate
	are shown here: Z=0~m (red circles), Z=$+$12~m (blue squares) and Z=$-$12~m (magenta triangles).
	The shaded area at the top of the plot indicates the neutron capture time constant in pure water
	\cite{Super-Kamiokande:2015xra}.
	Though the July 29 and August 5 data points at Z=$+$12~m are out of range (too high), they are nevertheless consistent with the pure water shaded region within about one sigma.
	}
	\label{fig:History_Of_GdCaptureTime}
\end{figure}

The neutron thermalization and capture time constants were considered when fitting the event candidate time distribution, as well as the presence of background events which were uniformly distributed in time.
For the conversion from capture lifetime to Gd concentration, a Geant4 Monte Carlo simulation 
was applied (Fig.~\ref{fig:CaptureT_Gd_MC}).
More specifically, the Geant4.9.6p04 Monte Carlo simulation with
its high precision and thermal scattering models was used to simulate the thermalization, capture, and other processes involving neutrons in the Gd-loaded water.
G4NDL4.2 is applied as the neutron cross section library for this simulation.
Interactions with the stainless steel Am/Be source container, the BGO scintillator, and the acrylic case which encloses the Am/Be+BGO components are also taken into account (Fig.~\ref{fig:AmBe_source}).
The neutron capture time constant was reasonably stable,   within statistical errors, as shown by data taken between  September and November 2020 (Fig.~\ref{fig:History_Of_GdCaptureTime}); it was also stable at several depths within SK. 
By applying Gaussian fitting to the 21 data points taken during this period, a mean neutron capture lifetime of 115~$\pm$~1~(sys.+stat.)~$\mu$s and a Gd concentration of 111~$\pm$~2~(sys.+stat.)~ppm were obtained.
The Gd concentration obtained by the Am/Be measurement is consistent with the estimation in Section~\ref{S:4-3}.



\section{Conclusion}
In the summer of 2020, 13.2 tons of Gd$_2$(SO$_4$)$_3$ $\cdot$8H$_2$O was dissolved into Super-Kamiokande and in so doing  a large-scale (50 kiloton) gadolinium-enhanced water Cherenkov detector was realized for the first time in the world.
During this Gd loading, laminar flow of water in the tank was successfully achieved through careful control the Gd concentration and the water temperature. The Gd concentration throughout the tank became homogeneous shortly after the addition of Gd was completed. Next, the transparency of the Gd-loaded water improved after starting water recirculation with the new SK water system. Following a three month post-loading commissioning phase, the Gd-enhanced water's transparency in SK became as good as that in the experiment's previous pure water phases, demonstrating that the new SK-Gd water system works effectively for the purification of 0.021$\%$ Gd$_2$(SO$_4$)$_3$ water.

Based on this success, increasing the Gd concentration is planned for the next SK phase, which is currently expected to begin in 2022. The next target concentration is 0.06$\%$ Gd$_2$(SO$_4$)$_3$ or 0.03$\%$ Gd, which should result in 75\% neutron tagging efficiency. 
Having gadolinium in the Super-Kamiokande water will enable the detector to positively identify low energy antineutrinos, allowing it to make the world's first observation of the diffuse supernova neutrino background flux via inverse beta decay.


\section*{Acknowledgment}
We gratefully acknowledge the cooperation of the Kamioka Mining and Smelting Company.
The Super‐Kamiokande experiment has been built and operated from funding by the 
Japanese Ministry of Education, Culture, Sports, Science and Technology, the U.S.
Department of Energy, and the U.S. National Science Foundation. Some of us have been 
supported by funds from the National Research Foundation of Korea NRF‐2009‐0083526
(KNRC) funded by the Ministry of Science, ICT, and Future Planning and the Ministry of
Education (2018R1D1A3B07050696, 2018R1D1A1B07049158), 
the Japan Society for the Promotion of Science (JSPS KAKENHI Grant Numbers JP19H05807,JP26000003), the National
Natural Science Foundation of China under Grants No.11620101004, the Spanish Ministry of Science, 
Universities and Innovation (grant PGC2018-099388-B-I00), the Natural Sciences and 
Engineering Research Council (NSERC) of Canada, the Scinet and Westgrid consortia of
Compute Canada, the National Science Centre, Poland (2015/18/E/ST2/00758),
the Science and Technology Facilities Council (STFC) and GridPPP, UK, the European Union's 
Horizon 2020 Research and Innovation Programme under the Marie Sklodowska-Curie grant
agreement no.754496, H2020-MSCA-RISE-2018 JENNIFER2 grant agreement no.822070, and 
H2020-MSCA-RISE-2019 SK2HK grant agreement no. 872549.

\clearpage
\appendix
\section{Specifications of SK-Gd Water System}
\label{Append:1}
\subsection{Gd Dissolving System}
%
%
\begin{itemize}
    \item{\bf Measuring hopper} -- Size: 700~mm $\phi$, 900~mm H; material:  SUS304; inside surface finish: buffing \# 250; effective volume: 200~$\ell$.
    \item {\bf Circle feeder} -- Maximum supply: 59.5~kg/min; minimum supply: 22.3~kg/min (at the factory test); rotation speed: 5.49-0.55~rpm; material: SUS304 (powder contact components w/ Teflon coating); motor: 1.5~kW 4 pole.
    \item {\bf Shear blender} -- The shear blender consists of three parts:
    \begin{itemize}
        \item{\bf Dissolving hopper} -- Volume: 66 $\ell$; material: SUS316L.
        \item{\bf Self-priming sanitary pump} -- Model: SIPLA Adapta 28.1; flow rate: 23~m$^3$/h; head: 20~m; power: 7.5~kW (4 pole, 200~V); rotation speed: 1750~rpm.
        \item{\bf Shear pump} -- Model: EMP305; flow rate: 15~m$^3$/h; head: 10~m; power: 7.5~kW (2 pole, 200~V); rotation speed: 3500~rpm.
    \end{itemize}
    \item{\bf Solvent tank} -- Tank capacity: 6~m$^3$ (1922~mm $\phi$, 2350~mm H); material: PE; thickness: 9.5~mm.
    \item{\bf Dissolving tank} -- Tank capacity: 4~m$^3$ (1740~mm $\phi$, 1780~mm H (cylinder), 280~mm H (taper)); material: PE; thickness: 9~mm.
    \item{\bf Solution tank} -- Tank capacity: 6~m$^3$ (1922~mm $\phi$, 2350~mm H); material: PE; thickness: 9.5~mm.
    \item{\bf Pumps} -- Three pumps are in the dissolving system:
    \begin{itemize}
        \item{\bf Supply pump after solvent tank} -- Flow rate: 48~m$^3$/h; head: 20~m; power: 5.5~kW (2 pole, 200~V); speed: 3600~rpm. 
        \item {\bf Transfer pump after dissolving tank} -- Flow rate: 48~m$^3$/h; head: 20~m; power: 5.5~kW (2 pole, 200~V); speed: 3600~rpm. 
        \item {\bf Injection pump after solution tank} -- Flow rate: 12~m$^3$/h; head: 45~m; power: 3~kW (2 pole, 200~V); speed: 3600~rpm. 
    \end{itemize}
\end{itemize}

\subsection{Pretreatment System}


\begin{itemize}
    \item {\bf Prefilter} -- Nominal pore size: 1~$\mu$m; size (one module): 62~mm $\phi$ outer, 30~mm $\phi$ inner, and 750~mm length;  material: polypropylene; number of modules: 6.
    \item {\bf TOC lamp (UV oxidation)} -- Wavelengths 253.7~nm and 184.9~nm; power: 0.81~W; Chiyoda Kohan Steritron WOX (lamp: CX1501).
    \item {\bf Cation exchange resin tank} -- Size: 1600~mm $\phi$ 1460  H; material: SUS304; resin volume: 1200~$\ell$; resin type: AMBERJET 1020(Gd).
    \item {\bf Anion exchange resin tank} -- Size: 1600mm $\phi$ 1460mm H; material: SUS304; resin volume: 2400~$\ell$; resin type: AMBERJET 4400(SO$_4$).
    \item {\bf Middle filter} -- Nominal pore size: 1~$\mu$m; size (one module): 62~mm $\phi$ outer, 30~mm $\phi$ inner, and 750~mm length;  material: polypropylene;  number of modules: 6.
    \item {\bf UV sterilizer} --  Wavelength 253.7~nm; power: 0.3~W; Chiyoda Kohan Steritron UEX (lamp: CS1001N).
    \item {\bf Postfilter} -- Nominal pore size: 0.2~$\mu$m; size (one module): 62~mm $\phi$ outer, 30~mm $\phi$ inner, and 750~mm length;  material: polypropylene;  number of modules: 6.
\end{itemize}
\subsection{Water Recirculation System}


\begin{itemize}
    \item {\bf Return water filter} -- Nominal pore size: 1~$\mu$m; size (one module): 62~mm $\phi$ outer, 30~mm $\phi$ inner, and 750~mm length;  material: polypropylene;  number of modules: 40.
    \item {\bf First buffer tank} -- Tank capacity: 10~m$^3$ (2280~mm $\phi$, 2780~mm H); material: PE; thickness: 12.5~mm.
    \item {\bf Heat exchange unit (HE) after the relay pump} -- Heat transfer area: 15.80~m$^2$; plate material: SUS316 (electrolytic polishing finish); plate gasket: EPDM with PTFE coating.
    \item {\bf TOC lamp (UV oxidation)} -- Wavelengths: 253.7~nm and 184.9~nm; power: 4.02~W; Chiyoda Kohan Steritron WOX (lamp: CX1501).
    \item {\bf Cation exchange resin tank} -- Size: 2100~mm $\phi$ 1610~mm H; material: SUS304; resin volume: 2400~$\ell$; resin type: AMBERJET1020(Gd).
    \item {\bf Anion exchange resin tank} -- Size: 2100~mm $\phi$ 1610~mm H; material: SUS304; resin volume: 4600~$\ell$; resin type: AMBERJET4400(SO$_4$). 
    \item {\bf Middle filter} -- Nominal pore size: 1~$\mu$m; size (one module): 62~mm $\phi$ outer, 30~mm $\phi$ inner, and 750~mm length;  material: polypropylene; number of modules: 40.
    \item {\bf UV sterilizer} --  Wavelength: 253.7~nm; power: 0.97~W; Chiyoda Kohan Steritron UEX (lamp: CS1001N).
    \item {\bf Ultrafiltration modules (UF)} -- Nitto NTU-3306-K6R;  inner/outer diameter of the capillary membrane:  0.7~mm / 1.3~mm; effective membrane area: 30~m$^2$/module; number of modules in one unit: 12; molecular weight cut-off: 6000; processed water TOC: $\leq$ 5 ppb; material:  polysulfone (capillary membrane), polysulfone (housing).
    \item {\bf Second buffer tank} -- Tank capacity: 20~m$^3$ (2710~mm $\phi$, 3810~mm H); material: PE; thickness: 15~mm.
    \item {\bf Heat exchange unit after the supply pump} -- Heat transfer area: 24.74~m$^2$; plate material: SUS316 (electrolytic polishing finish); plate gasket:  EPDM with PTFE coating.
    \item {\bf Membrane degasifier (MD)} -- DIC SEPAREL\textregistered EF-040P-JO; module size: 180~mm $\phi$ 673~mm H; number of modules: 60; operation pressure: $-92$~kPa; purge gas: radon-free air~\cite{Fukuda:2003skdet, Nakano:2017rsy}; purge gas flow rate: 1~$\ell$/min (per module).
    \item {\bf Final Heat exchange unit} -- Heat transfer area: 21.15~m$^2$, plate material: SUS316; plate gasket: butyl (isobuthene-isoprene) rubber.
     \item {\bf Pumps} --  Pumps in the recirculation system:
   \begin{itemize} 
        \item {\bf Return pump after Super-K tank} -- Flow rate: 60~m$^3$/h; head: 70~m; power: 36~kW (200~V); speed: 3600~rpm. 
        \item {\bf Relay pump after first buffer tank} -- Flow rate:  60~m$^3$/h; head: 40~m; power: 18~kW (200~V); speed: 3600~rpm. 
        \item {\bf Supply pump after second buffer tank} -- Flow rate: 60~m$^3$/h; head: 62~m; power: 22~kW (200~V); speed 3600~rpm. 
  \end{itemize}
\end{itemize}

\section{Water Flow in the SK Tank}
\label{Append:2}
Precise control of the water flow in the Super-Kamiokande tank is important not only for gadolinium loading, but also for maximizing the physics performance of the detector. Such precise control plays critical roles in reducing radioactive backgrounds in the detector's fiducial volume as well as in improving water transparency. In this section, the operation and modeling of the water flow in the SK detector are described in more detail.

\subsection{Water Piping}
\begin{figure}[htb]
\centering\includegraphics[width=0.6\linewidth]{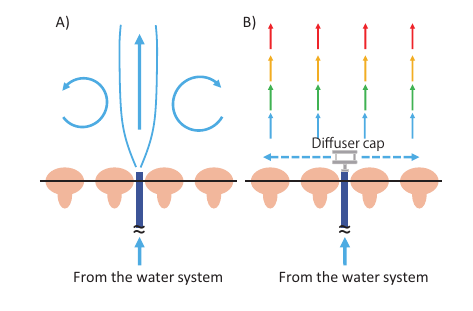}
\caption{Illustration of water flow near the ID bottom outlets A) without, and B) with the diffuser caps installed in February 2020.}
\label{fig:diffuser_cap}
\end{figure}

As illustrated in Fig.~\ref{fig:sk_water_flow_dissolving}, there are six groups of water inlets/outlets in the SK tank. The total flow and flow direction for each group can be individually adjusted by operating a set of valves located on top of the tank. 
These water pipes were upgraded during the refurbishment of the Super-Kamiokande detector in 2018--2019 to allow more precise control of water flow and also to permit a higher maximum total 
flow rate of 120~m$^3$/h (previously 60~m$^3$/h).
In addition, diffuser caps, illustrated in Fig.~\ref{fig:diffuser_cap}, were installed at all ID bottom and OD annular outlets in order to minimize the zones of turbulence near the water ejection points. 
It should be kept in mind that the separation between the ID and OD regions is not watertight and there potentially exists water flow between them. 

\subsection{Water Flow Model}

Typically, water is injected into the bottom region of the tank and extracted from the top. 
Once inside the tank, this injected water gradually becomes warmer due to heat produced by PMTs and the magnetic field compensation coils\footnote{There could also be heat transfer from the rock surrounding the detector tank, whose contribution is unknown. However, it was found that temperature change of water in the SK detector is consistent with the sum of the expected impacts from the PMTs and the coil.}; this heating results in a positive water temperature gradient from the bottom to the top of the detector. 

\begin{figure}[htb]
\begin{center}
\includegraphics[width=0.45\textwidth]{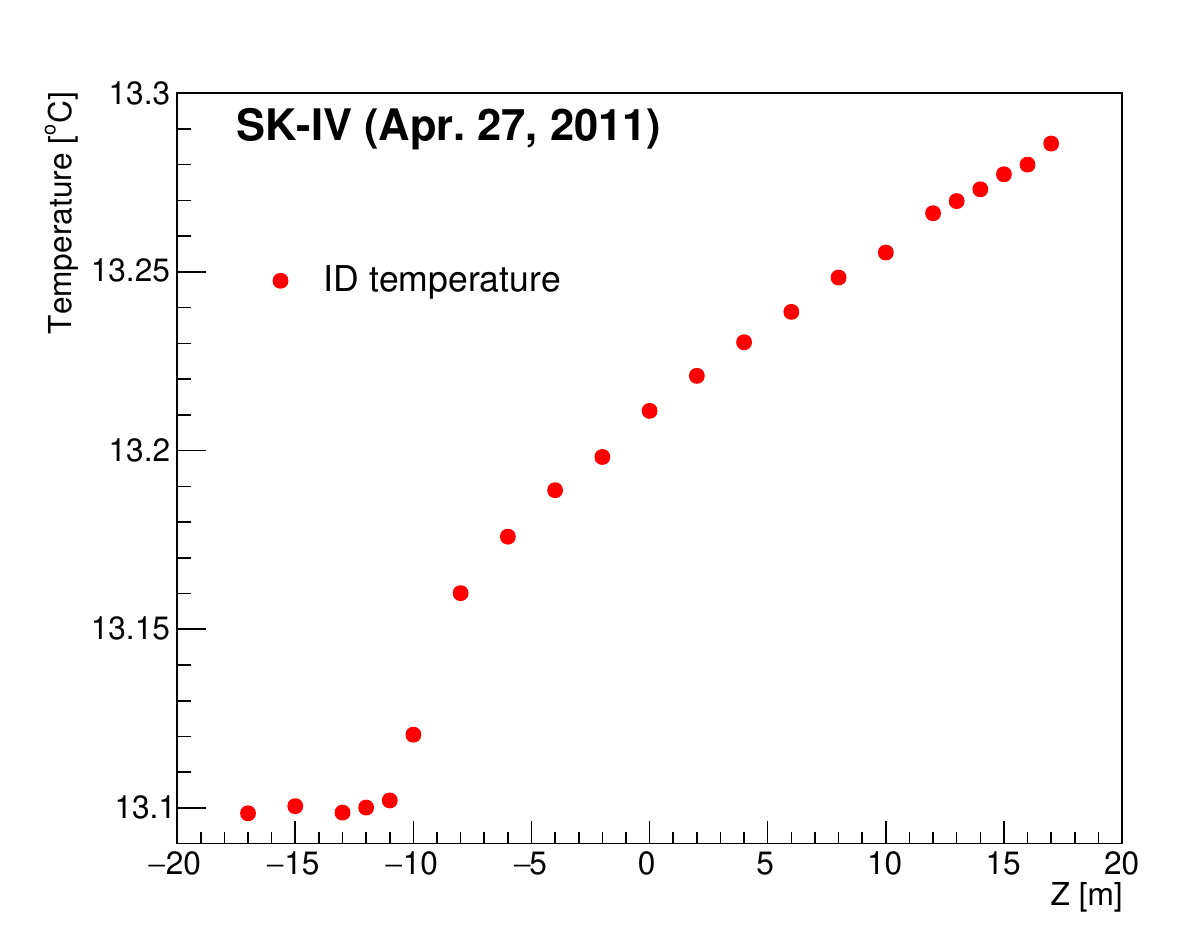}
\includegraphics[width=0.45\textwidth]{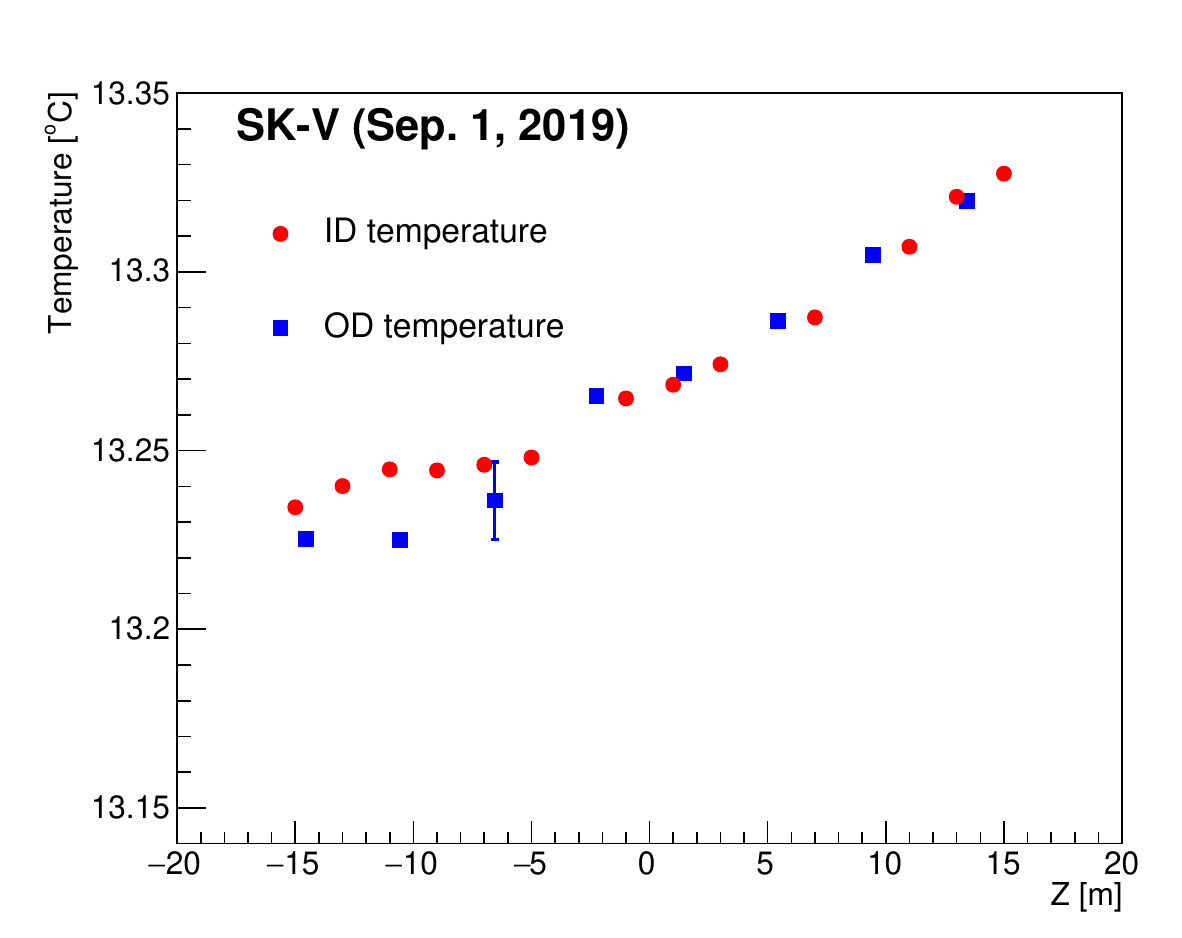}
\caption{
  Typical water temperature distributions in the SK tank
  during the SK-IV (left) and SK-V (right) periods.
  The left figure was taken from Ref.~\cite{Abe:2016nxk}.
}
\label{fig:SKtemp}
\end{center}
\end{figure}

Figure~\ref{fig:SKtemp} shows typical temperature profiles along the Z (vertical) coordinate of the detector during the SK-IV (2009-2018) and SK-V (2019-2020) phases.
The region at a roughly constant temperature (Z$ < \sim -10~\mathrm{m}$ for SK-IV)
indicates a convection zone, while the regions with constant slope as
a function of z indicate a steady vertical flow.
During the SK-V period there is almost no convection region inside the ID,
and the steady flow seems to be established at Z $> -5$~m.

\begin{figure}[htb]
\begin{center}
\includegraphics[width=0.48\textwidth]{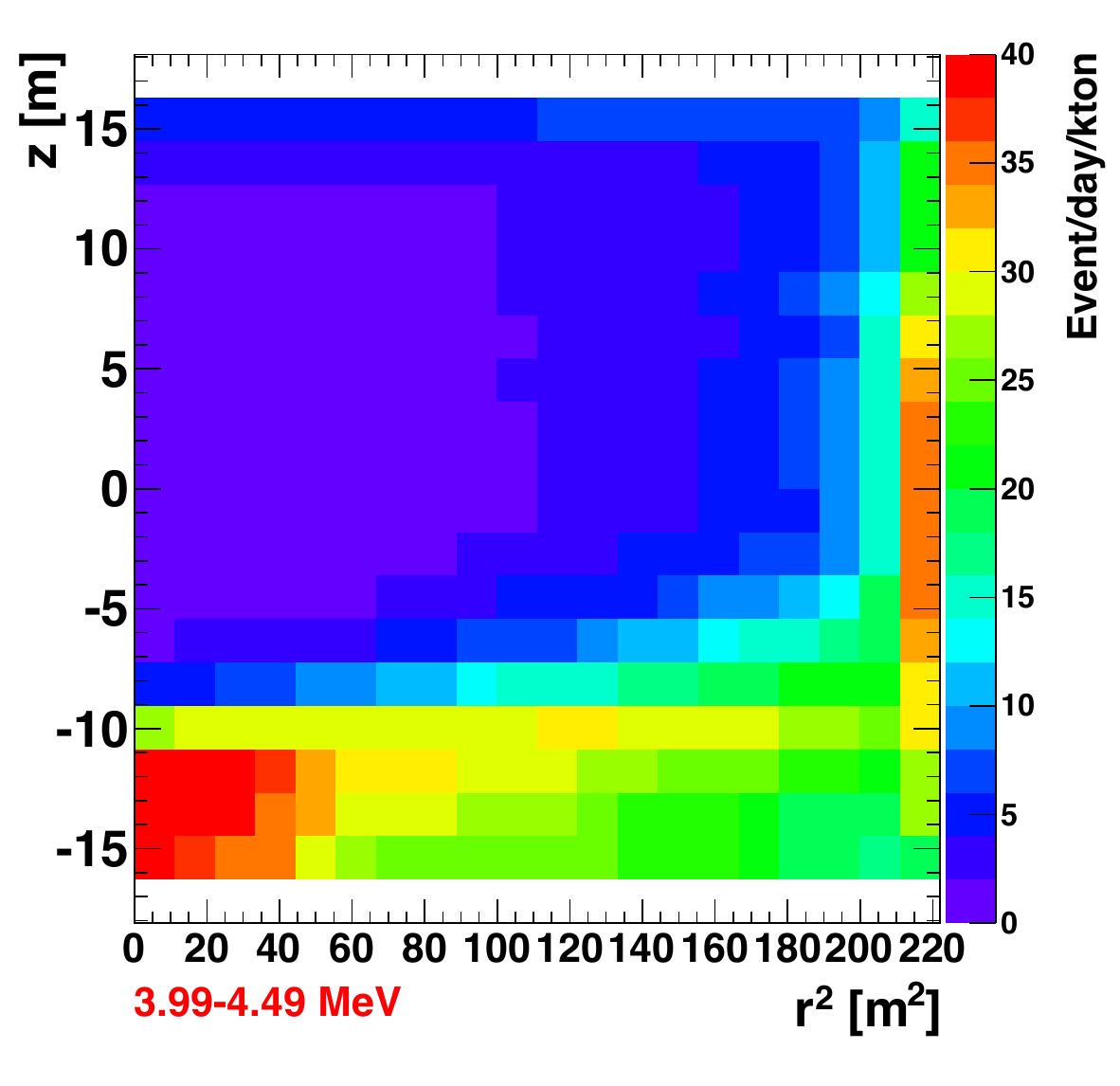}
\includegraphics[width=0.48\textwidth]{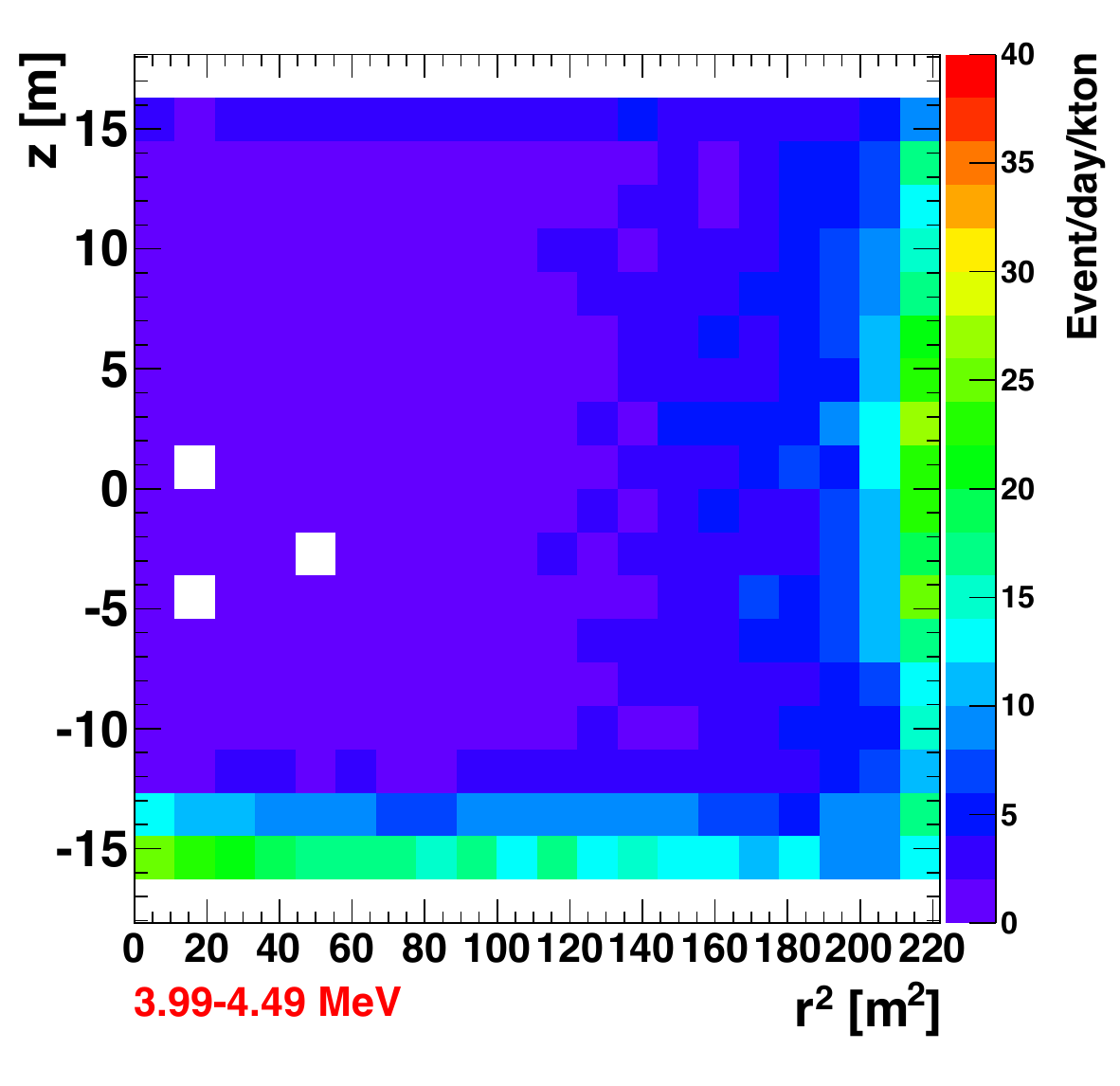}
\caption{Vertex distributions in z vs. r$^2$ for events in the range $3.99 < E_{\rm{kin}} < 4.49$~MeV
  for SK-IV (left) and SK-V (right).
  Selection cuts based on those used for the solar neutrino analysis of SK data~\cite{Abe:2016nxk}, except for the tight fiducial volume cut, are applied. 
  The livetime for the SK-IV sample is  2047 days, 
  while it is 107 days for the SK-V sample.
  Compared to the SK-IV configuration, the convection region (high event rate region) was significantly reduced in SK-V.
}
\label{fig:lowe_vertex}
\end{center}
\end{figure}

The picture described above is supported by evaluating low energy background rates -- events with a few MeV -- in the detector. The dominant source of the low energy events at Super-Kamiokande is radon (Rn) emanated from the detector components. 
This Rn dissolves into the SK water and is then carried to various locations by water flow.
If there is convection, Rn from the PMTs and their associated hardware is quickly distributed 
into the center part of the detector and thus increases low energy
backgrounds in the fiducial volume.  In contrast, having slow, steady flow from bottom to top
naturally minimizes such issues, in part because the Rn emanated from surfaces has more time to decay before reaching the fiducial volume.
Figure~\ref{fig:lowe_vertex} shows vertex distributions for events in the range of 
$3.99 < E_{\rm{kin}} < 4.49$~MeV for SK-IV and SK-V. 
Enhancement of the event rate in the convection region at Z $< -10$~m in SK-IV was clearly observed, and  
increased Rn concentration in the bottom convection region was directly observed by 
analyzing water sampled from the detector tank~\cite{Nakano:2019bnr}.
Such enhancement was largely eliminated in SK-V, in which convection was significantly suppressed by water flow tuning. This expanded the low background region in the center of the 
detector and improved the experiment's low energy solar neutrino measurements.

The size of the
convection region is largely determined by the water injection scheme. On the other hand, the flow in the steady flow
region is mostly governed by the heat produced in the tank and
is almost independent of the water injection pattern. 
In the steady flow region, the water temperature at a given Z position
is equivalent in the ID and OD (right panel of Figure~\ref{fig:SKtemp}), which indicates no 
major water flow between them.
Under these conditions, the temperature gradient along the Z direction,
$dT/dZ$, can be described as 
\begin{equation}
  \label{eq:1}
  \frac{dT}{dZ}\propto \frac{Q}{v_Z},
\end{equation}
where $Q$ is the heat injected to the system and $v_Z$ is the vertical
speed of the water flow in the tank. This relationship can be applied
to the
bulk behavior of the water, and also to the local behavior at a
given location in the tank.
In addition, $dT/dZ$ is approximately a constant across the
steady flow region.
Therefore, we can deduce a simple relation that the vertical flow
speed at a given position $v_Z(X,Y,Z)$ is proportional to the amount of
heat injected to the same position, $Q(X,Y,Z)$.

\begin{figure}[htb]
\begin{center}
\includegraphics[width=0.4\textwidth]{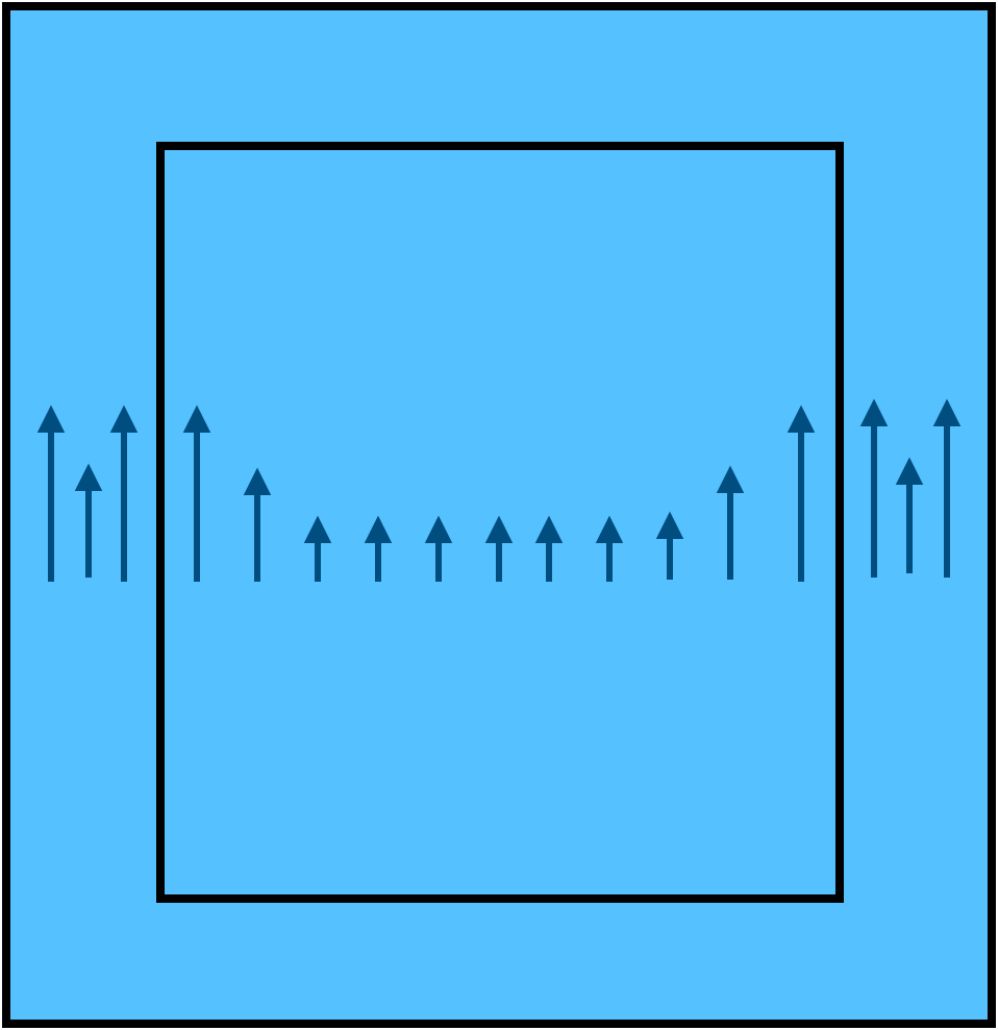}
\caption{A schematic of expected vertical water flow in the steady
  flow region in the SK tank.  Arrow lengths indicate relative velocities.
}
\label{fig:water_flow_cartoon}
\end{center}
\end{figure}

This relationship indicates that the water flow speed is not constant
across the detector volume, because the distribution of the heat sources
 (i.e. the PMTs and the magnetic field compensation coil) are
not uniform.
As illustrated in Figure~\ref{fig:water_flow_cartoon}, the flow is expected to be faster near the PMTs and the detector walls where the coils are mounted,
while it becomes much smaller near the detector center.
The slowness of the vertical flow in the central region was
confirmed by injecting Rn-rich water into the SK tank and tracing its spatial distribution over time.

This flow configuration works well for reducing Rn backgrounds in the
SK detector.
Most of the Rn-rich water near the PMT structure flows vertically at a
relatively fast speed, and then returns to the water purification system without
going through the central region of the detector.
The slowness of water replacement in the ID region also helps reduce
Rn backgrounds, as most of the 
Rn atoms decay before reaching the fiducial volume.
Additionally, it should be noted that long water attenuation lengths have been maintained 
with this configuration, typically $\sim$90~m, based on measurements with cosmic-ray muons.

\subsection{Water Flow for Gd Loading}
\label{SS:B3}
Although this steady water flow is ideal for studying low energy neutrinos, it was not ideal for Gd loading as this flow does not efficiently replace water, especially near the central part of the detector.
Therefore, we lowered the supply water temperature at the same time we started injecting Gd, in order to create a larger density difference between the new Gd-loaded water and the pure water remaining in the tank. This density difference ensured the efficient replacement of water throughout the detector volume as described in Section~\ref{sec:water_flow}. 

\begin{figure}[hbt!]
\begin{center}
\includegraphics[width=0.9\textwidth]{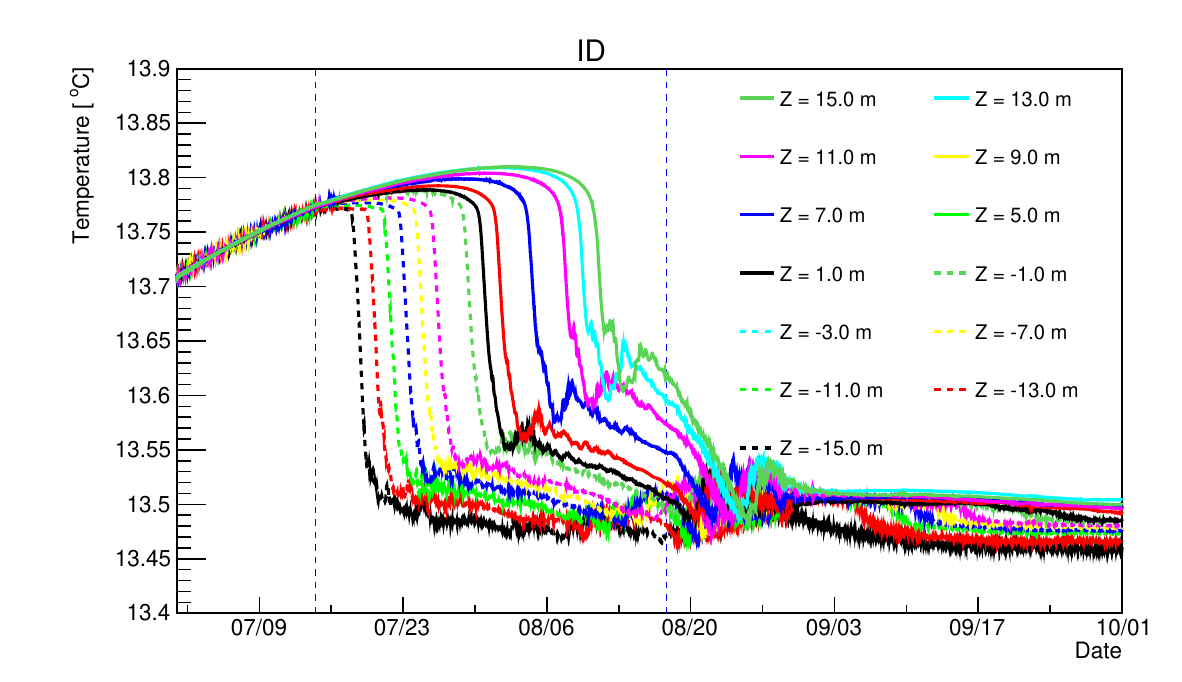}
\includegraphics[width=0.9\textwidth]{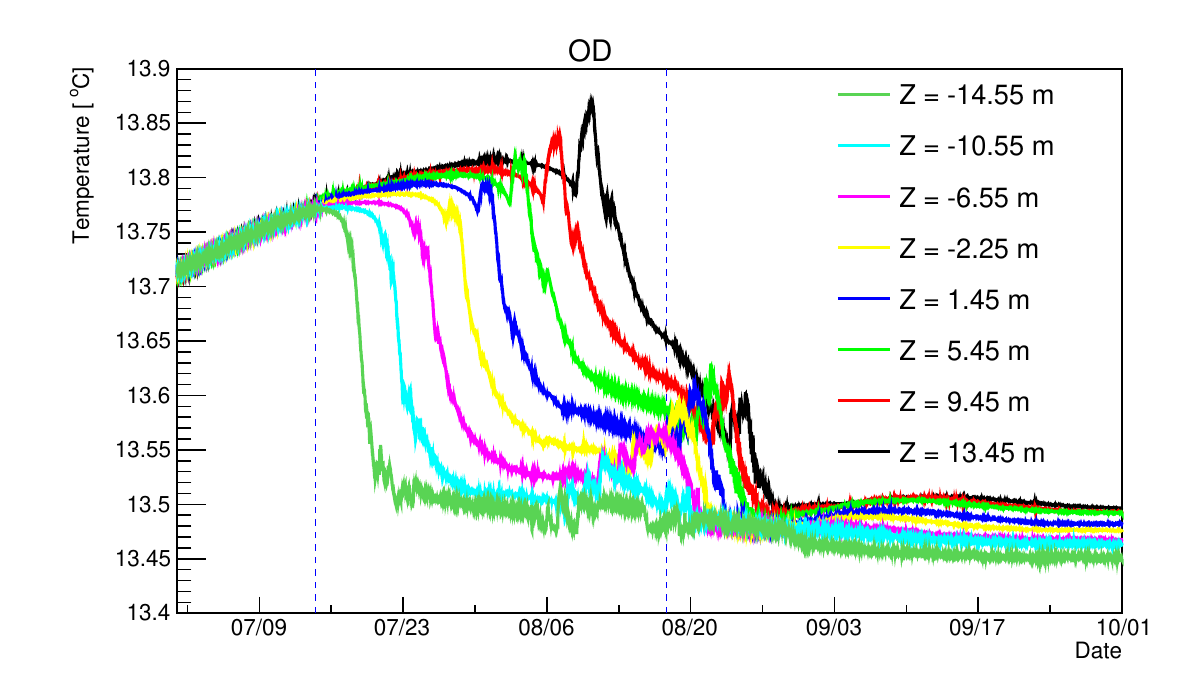}
\caption{Time evolution of water temperature around the 2020 Gd loading period
at different vertical positions in the SK inner detector (top) and outer detector (bottom).
Vertical dashed lines indicate the start and end dates of Gd loading.
Relative biases of the temperature sensors were corrected using data taken between July 1, 2020, and July 10, 2020, during which time the water in the tank was in a state of full convection.
}
\label{fig:temp_profile_loading}
\end{center}
\end{figure}

Movement of the Gd-loaded water in the tank was tracked by monitoring water temperature. Figure~\ref{fig:temp_profile_loading} shows the time evolution of water temperature at different locations around the period of Gd loading. Arrival of Gd-loaded water at a given location in the detector was clearly observed with a sharp drop of water temperature there. This demonstrates that Gd-loaded water successfully replaced pure water from the bottom, and that the replacement uniformly occurred in both ID and OD regions.
Smaller fluctuations of water temperature after the sharp drops were observed, which indicate that there was some turbulence within the Gd-water region. After recirculating water for another month after the conclusion of Gd loading, the temperature fluctuations naturally faded away and a stable detector condition was realized.



\newpage
\bibliographystyle{elsarticle-num-names}
\bibliography{main.bib}







\end{document}

%% file: Authors.tex
\address[AFFicrr]{Kamioka Observatory, Institute for Cosmic Ray Research, University of Tokyo, Kamioka, Gifu 506-1205, Japan}
\address[AFFkashiwa]{Research Center for Cosmic Neutrinos, Institute for Cosmic Ray Research, University of Tokyo, Kashiwa, Chiba 277-8582, Japan}
\address[AFFicrronly]{Institute for Cosmic Ray Research, University of Tokyo, Kashiwa, Chiba 277-8582, Japan}
\address[AFFmad]{Department of Theoretical Physics, University Autonoma Madrid, 28049 Madrid, Spain}
\address[AFFbcit]{Department of Physics, British Columbia Institute of Technology, Burnaby, BC, V5G 3H2, Canada }
\address[AFFubc]{Department of Physics and Astronomy, University of British Columbia, Vancouver, BC, V6T1Z4, Canada}
\address[AFFbu]{Department of Physics, Boston University, Boston, MA 02215, USA}
\address[AFFuci]{Department of Physics and Astronomy, University of California, Irvine, Irvine, CA 92697-4575, USA}
\address[AFFcsu]{Department of Physics, California State University, Dominguez Hills, Carson, CA 90747, USA}
\address[AFFcnm]{Institute for Universe and Elementary Particles, Chonnam National University, Gwangju 61186, Korea}
\address[AFFduke]{Department of Physics, Duke University, Durham NC 27708, USA}
\address[AFFllr]{Ecole Polytechnique, IN2P3-CNRS, Laboratoire Leprince-Ringuet, F-91120 Palaiseau, France}
\address[AFFfukuoka]{Junior College, Fukuoka Institute of Technology, Fukuoka, Fukuoka 811-0295, Japan}
\address[AFFgifu]{Department of Physics, Gifu University, Gifu, Gifu 501-1193, Japan}
\address[AFFgist]{GIST College, Gwangju Institute of Science and Technology, Gwangju 500-712, Korea}
\address[AFFuh]{Department of Physics and Astronomy, University of Hawaii, Honolulu, HI 96822, USA}
\address[AFFicl]{Department of Physics, Imperial College London , London, SW7 2AZ, United Kingdom}
\address[AFFifirse]{Institute For Interdisciplinary Research in Science and Education, Quy Nhon 55121, Binh Dinh, Vietnam. }
\address[AFFbari]{Dipartimento Interuniversitario di Fisica, INFN Sezione di Bari and Universit\`a e Politecnico di Bari, I-70125, Bari, Italy}
\address[AFFnapoli]{Dipartimento di Fisica, INFN Sezione di Napoli and Universit\`a di Napoli, I-80126, Napoli, Italy}
\address[AFFpadova]{Dipartimento di Fisica, INFN Sezione di Padova and Universit\`a di Padova, I-35131, Padova, Italy}
\address[AFFroma]{INFN Sezione di Roma and Universit\`a di Roma ``La Sapienza'', I-00185, Roma, Italy}
\address[AFFkeio]{Department of Physics, Keio University, Yokohama, Kanagawa, 223-8522, Japan}
\address[AFFkek]{High Energy Accelerator Research Organization (KEK), Tsukuba, Ibaraki 305-0801, Japan}
\address[AFFkcl]{Department of Physics, King's College London, London, WC2R 2LS, UK }
\address[AFFkobe]{Department of Physics, Kobe University, Kobe, Hyogo 657-8501, Japan}
\address[AFFkyoto]{Department of Physics, Kyoto University, Kyoto, Kyoto 606-8502, Japan}
\address[AFFliv]{Department of Physics, University of Liverpool, Liverpool, L69 7ZE, United Kingdom}
\address[AFFmiyagi]{Department of Physics, Miyagi University of Education, Sendai, Miyagi 980-0845, Japan}
\address[AFFnagoya]{Institute for Space-Earth Environmental Research, Nagoya University, Nagoya, Aichi 464-8602, Japan}
\address[AFFkmi]{Kobayashi-Maskawa Institute for the Origin of Particles and the Universe, Nagoya University, Nagoya, Aichi 464-8602, Japan}
\address[AFFpol]{National Centre For Nuclear Research, 02-093 Warsaw, Poland}
\address[AFFsuny]{Department of Physics and Astronomy, State University of New York at Stony Brook, NY 11794-3800, USA}
\address[AFFokayama]{Department of Physics, Okayama University, Okayama, Okayama 700-8530, Japan}
\address[AFFox]{Department of Physics, Oxford University, Oxford, OX1 3PU, United Kingdom}
\address[AFFral]{Rutherford Appleton Laboratory, Harwell, Oxford, OX11 0QX, UK }
\address[AFFseoul]{Department of Physics, Seoul National University, Seoul 151-742, Korea}
\address[AFFsheff]{Department of Physics and Astronomy, University of Sheffield, S3 7RH, Sheffield, United Kingdom}
\address[AFFshizuokasc]{Department of Informatics in Social Welfare, Shizuoka University of Welfare, Yaizu, Shizuoka, 425-8611, Japan}
\address[AFFstfc]{STFC, Rutherford Appleton Laboratory, Harwell Oxford, and Daresbury Laboratory, Warrington, OX11 0QX, United Kingdom}
\address[AFFskk]{Department of Physics, Sungkyunkwan University, Suwon 440-746, Korea}
\address[AFFtohoku]{Department of Physics, Tohoku University, Aoba, Sendai 9808578, Japan}
\address[AFFtokai]{Department of Physics, Tokai University, Hiratsuka, Kanagawa 259-1292, Japan}
\address[AFFtokyo]{The University of Tokyo, Bunkyo, Tokyo 113-0033, Japan}
\address[AFFtodai]{Department of Physics, University of Tokyo, Bunkyo, Tokyo 113-0033, Japan}
\address[AFFipmu]{Kavli Institute for the Physics and Mathematics of the Universe (WPI), The University of Tokyo Institutes for Advanced Study, University of Tokyo, Kashiwa, Chiba 277-8583, Japan}
\address[AFFtit]{Department of Physics,Tokyo Institute of Technology, Meguro, Tokyo 152-8551, Japan}
\address[AFFtus]{Department of Physics, Faculty of Science and Technology, Tokyo University of Science, Noda, Chiba 278-8510, Japan}
\address[AFFtoronto]{Department of Physics, University of Toronto, ON, M5S 1A7, Canada }
\address[AFFtriumf]{TRIUMF, 4004 Wesbrook Mall, Vancouver, BC, V6T2A3, Canada }
\address[AFFtsinghua]{Department of Engineering Physics, Tsinghua University, Beijing, 100084, China}
\address[AFFwu]{Faculty of Physics, University of Warsaw, Warsaw, 02-093, Poland}
\address[AFFwarwick]{Department of Physics, University of Warwick, Coventry, CV4 7AL, UK }
\address[AFFwinnipeg]{Department of Physics, University of Winnipeg, MB R3J 3L8, Canada }
\address[AFFynu]{Department of Physics, Yokohama National University, Yokohama, Kanagawa, 240-8501, Japan}


\author[AFFicrr,AFFipmu]{K.~Abe}
\author[AFFicrr]{C.~Bronner}
\author[AFFicrr,AFFipmu]{Y.~Hayato}
\author[AFFicrr]{K.~Hiraide}
\author[AFFicrr,AFFipmu]{M.~Ikeda}
\author[AFFicrr]{S.~Imaizumi}
\author[AFFicrr,AFFipmu]{J.~Kameda}
\author[AFFicrr]{Y.~Kanemura}
\author[AFFicrr]{Y.~Kataoka}
\author[AFFicrr]{S.~Miki}
% ...

\author{\\
(The Super-Kamiokande Collaboration)
}